\documentclass{WileyMSP-template}

\usepackage[utf8]{inputenc}
\usepackage[T1]{fontenc}

\usepackage{graphicx}
\usepackage{amsmath}
\usepackage{amssymb}
\usepackage{afterpage}
\usepackage{amsfonts}
\usepackage{braket}
\usepackage{float}
\usepackage{bm}
\usepackage{siunitx}
\usepackage{lipsum}
\usepackage[]{xcolor}
\usepackage[english]{babel}
\usepackage{hyperref}
\usepackage{physics}
\hypersetup{
	colorlinks,
	linkcolor={blue!90!black},
	citecolor={blue!90!black},
	urlcolor=	{blue!90!black}
}
\usepackage[]{placeins}
\usepackage{makecell}
\usepackage[sort,compress]{cite}

\begin{document}

\justify

\pagestyle{fancy}
\rhead{\includegraphics[width=2.5cm]{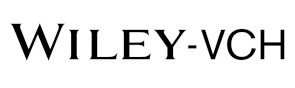}}

\title{Phonon-induced Markovian and non-Markovian effects on absorption spectra of moiré excitons in twisted transition metal dichalcogenide bilayers}

\maketitle


\author{Daniel Groll*}
\author{Anton Plonka}
\author{Kevin J\"urgens}
\author{Daniel Wigger}
\author{Tilmann Kuhn}

\begin{affiliations}
Dr. Daniel Groll, Anton Plonka, Dr. Kevin J\"urgens, Prof. Tilmann Kuhn\\
Institute of Solid State Theory, University of M\"unster, 48149 M\"unster, Germany\\
Email address: daniel.groll@uni-muenster.de

Dr. Daniel Wigger\\
Department of Physics, University of M\"unster, 48149 M\"unster, Germany

\end{affiliations}



\begin{abstract}

The properties of moiré excitons in twisted bilayers of transition metal dichalcogenides (TMDCs) vary significantly with the twist angle, ranging from quasi localized excitons with flat dispersions for small twist angles to delocalized excitons for larger ones. This twist angle dependence directly impacts the exciton-pho\-non coupling, which plays a significant role for the optical properties of these materials. In this work we theoretically investigate the twist angle dependent influence of pho\-nons on absorption spectra of intra\-layer moiré excitons in a twisted TMDC hetero-bilayer. For the lowest-lying intra\-layer moiré exciton we find that the exciton-pho\-non coupling interpolates between two physically distinct regimes when tuning the twist angle. At small twist angles non-Markov\-ian polarization dynamics and pho\-non sidebands dominate the properties of absorption spectra for localized excitons. For larger twist angles Markov\-ian processes become more important leading to additional line broadening. Furthermore, the absorption spectra here show a characteristic asymmetric peak similar to monolayer TMDCs. When taking into account multiple bright moiré exciton bands we find that intra\-band scattering due to optical pho\-nons has a significant impact on absorption spectra, effectively suppressing absorption peaks of higher lying bands when their bandwidth surpasses the optical pho\-non energy.
\end{abstract}


\twocolumn
\section{Introduction}\label{sec:intro}
In recent decades, material processing has paved the way for modern research that focuses on systems with nanoscale sizes and reduced dimensionalities~\cite{chemla1993optics, geim2007rise,brus2014size, mueller2018exciton}. Such nano-structuring provides a confinement for charge carriers and thereby has a strong impact on the optical properties. In modern semiconductor nano-optics, in particular 2-dimensional (2D) van der Waals materials and 0D single-photon emitters play an outstanding role, because of their perspective for future optoelectronic and quantum technological applications~\cite{mueller2018exciton,heindel2023quantum, montblanch2023layered,niehues2024excitons}.

For many solid state light sources the unavoidable interplay between charge and lattice excitations has a significant impact on the optical properties~\cite{krummheuer2002theory,ramsay2010damping,christiansen2017phonon,niehues2018strain,wigger2019phonon,brem2020phonon,preuss2022resonant}. To fully harness their potential, we therefore need to understand the influence of exciton-pho\-non interaction. It is known from localized 0D emitters, e.g., from quantum dots or color centers, that the interaction of excitons with pho\-nons manifests itself primarily in the formation of sidebands~\cite{besombes2001acoustic,peter2004phonon,wigger2019phonon,wigger2020acoustic,preuss2022resonant}. These features stem from non-Markov\-ian pho\-non-in\-duced dephasing dynamics of ex\-citons caused by linear pho\-non coup\-ling~\cite{krummheuer2002theory, nazir2016modelling,preuss2022resonant}. In contrast, non-linear pho\-non effects are typically required to explain phonon-induced spectral line-broadening, i.e., Markov\-ian dephasing, in these 0D systems~\cite{muljarov2004dephasing,machnikowski2006change}. Due to the continuous exciton energy spectrum of 2D materials, the influence of pho\-nons is significantly different. Here, scattering with pho\-nons results in relaxation processes that can have a large impact on transport phenomena~\cite{yuan2017exciton,rosati2021dark,wagner2021nonclassical}. While this leads to a more pronounced role of Markov\-ian dynamics and line broadening compared to 0D systems, pho\-non sidebands often result in asymmetric spectral lines in these materials~\cite{christiansen2017phonon,niehues2018strain,lengers2020theory}.

The possibility of creating hetero-structures of van der Waals semiconductors allows to engineer the excitonic confinement and to interpolate between the 0D and 2D regimes by twisting two monolayers with respect to each other and generating a moir\'e super-lattice~\cite{yu2017moire, huang2022excitons,mak2022semiconductor, herrera2025moire}. This super-lattice leads to a twist angle dependent localization potential for the excitons, allowing for the transition between the two limiting cases of localized excitons with flat dispersions for small twist angles and delocalized excitons for larger twist angles~\cite{brem2020tunable, knorr2022exciton, jurgens2024theory}.





In this work we consider the platform of a twisted TMDC hetero-bilayer to study the twist angle dependent exciton-pho\-non coupling and its impact on the optical absorption spectrum of intra\-layer excitons theoretically. We present a time-convolutionless (TCL) master equation approach for the polarization dynamics of the moiré excitons coupled to pho\-nons. This allows us to perform efficient numerical simulations and gain analytical insight into the pho\-non-in\-duced dynamics~\cite{breuer2002theory,lengers2020theory, jurgens2024theory}. We discuss the influence of pho\-nons on the optical absorption properties of the lowest-lying bright moiré exciton extensively, focusing on the separate effects of acoustic and optical pho\-nons. Finally, we investigate the impact of pho\-nons in a system with multiple optically active moiré bands.


\section{Modeling polarization dynamics and absorption spectra of moiré intra\-layer excitons}
In this section we describe the theory for calculating absorption spectra of intra\-layer excitons in twisted TMDC bilayers. We will apply this theory to the case of MoSe$_2$ intra\-layer excitons in a twisted MoSe$_2$/ WSe$_2$ hetero-bilayer. Since MoSe$_2$ is a direct bandgap semiconductor in the monolayer, we start with an electronic two-band model in effective mass approximation in the usual fashion (see App.~\ref{app:exciton_picture}), assuming sufficiently resonant excitation of the corresponding 1s exciton, as well as fixing the polarization of the exciting light field such that only excitons in the K valley of the monolayer are created~\cite{xiao2012coupled, wang2018colloquium}. Considering weak optical excitation to obtain the linear absorption spectra, we can assume low densities of electrons and holes, such that the \textit{homogeneous} intra\-layer excitons, i.e., the excitons in the homogeneous material without moiré structure, are described by the Hamiltonian~\cite{katsch2018theory, lengers2020theory}
\begin{equation}\label{eq:ex-hom}
	H_{\rm ex-hom}=\sum_{\bm{K}} E_{\bm{K}}Y_{\bm{K}}^{\dagger} Y_{\bm{K}}^{}\,.
\end{equation}
Here we restrict ourselves to the lowest-lying 1s exciton, having the largest oscillator strength~\cite{chernikov2014exciton}, with $E_{\bm{K}}=E_{0}+\frac{\hbar^2\bm{K}^2}{2M}$ being the dispersion relation, $M$ the total exciton mass and $E_0$ the energy of the bright 1s exciton, i.e., the one with vanishing momentum $\hbar\bm{K}=\bm{0}$. The \textit{homogeneous} exciton annihilation $Y_{\bm{K}}^{}$ and creation operators $Y_{\bm{K}}^{\dagger}$ fulfill bosonic commutation relations due to the assumed low-densi\-ty limit for electrons and holes~\cite{katsch2018theory}.

Due to the two-band model the theory is, strictly speaking, limited to direct semiconductors like MoSe$_2$ or MoS$_2$. It is well-known that in the indirect W-based TMDCs intervalley scattering between bright and momentum-dark excitons plays a crucial role, especially for photoluminescence spectra~\cite{selig2016excitonic,selig2018dark,brem2020phonon}. 



\subsection{Influence of the moiré potential}
\begin{figure}
	\centering
	\includegraphics[width=\linewidth]{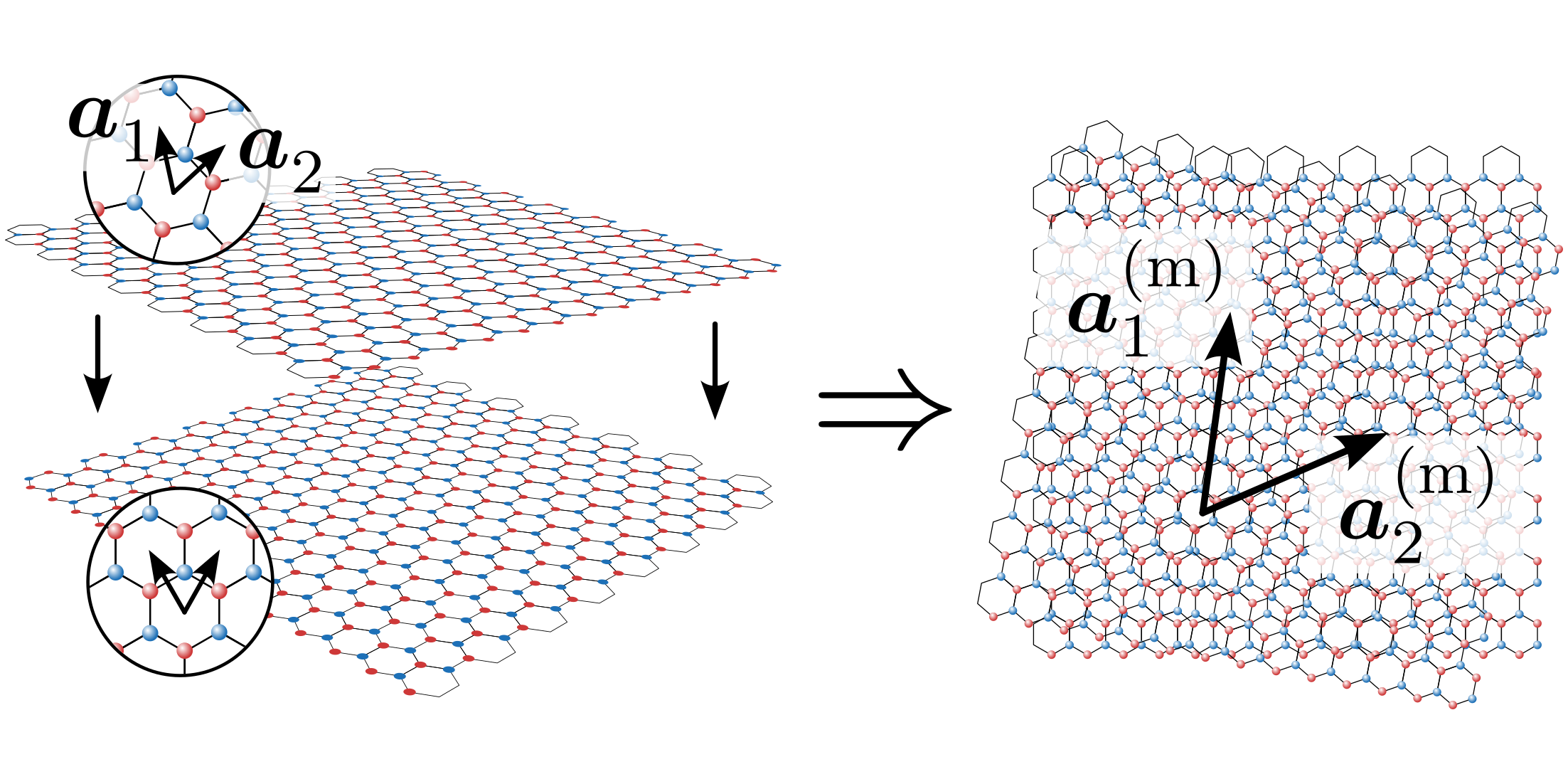}
	\caption{Schematic picture of the moir\'e lattice generation. The two sets of monolayer lattice basis vectors (left) and the moir\'e superlattice basis vectors (right) are respectively indicated by two vectors sharing the same origin.}
	\label{fig:moire_lattice}
\end{figure}
Stacking two TMDC monolayers on top of each other at a twist angle $\theta$, with $\theta=\mathcal{O}(1^{\circ})$ in the following, leads to a moiré superlattice as depicted in Fig.~\ref{fig:moire_lattice}. The moiré superlattice basis vectors $\bm{a}_{j=1,2}^{(\rm m)}$ can be determined from the basis vectors $\bm{a}_{j=1,2}$ of the monolayers, assumed to be approximately identical for both layers, via~\cite{nam2017lattice}
\begin{equation}\label{eq:moire_basis}
	\bm{a}_j^{(\rm m)}=\frac{e^{i\theta/2}}{2i\sin(\theta/2)}\bm{a}_j\,.
\end{equation}
Here the two-dimensional vectors have been represented by complex numbers in the two-dimensional complex plane $\bm{a}\hat{=}a_x+ia_y$, such that a rotation by the angle $\theta$ can be written as a multiplication by $e^{i\theta}$. From Eq.~\eqref{eq:moire_basis} we can see that the length of the moiré basis vectors changes with the twist angle. Their length and thus the size of the superlattice unit cell generally decreases with increasing twist angle for the twist angles $\theta=\mathcal{O}(1^{\circ})$ relevant for this work. This implies that the size of the first moiré Brillouin zone (MBZ) of the reciprocal superlattice shrinks with decreasing twist angle. For twist angles $\theta=\mathcal{O}(1^{\circ})$ the MBZ is much smaller than the conventional Brillouin zone of the monolayers. Note that the direct and reciprocal moiré superlattice inherit the hexagonal honeycomb structure from the monolayers via Eq.~\eqref{eq:moire_basis}. 

The presence of the moiré superlattice generates a potential for electrons and holes that is periodic with respect to translations by integer multiples of the moiré basis vectors in Eq.~\eqref{eq:moire_basis}. This leads to an effective potential for the excitons~\cite{brem2020tunable, knorr2022exciton}
\begin{equation}\label{eq:V_m}
	V_{\rm m}=\sum_{{\bm K},{\bm G}} V_{\bm{G}} Y^{\dagger}_{\bm{K}+\bm{G}}Y^{}_{\bm K}\,,
\end{equation}
describing scattering processes in which the homogeneous exciton changes its wave vector by a reciprocal moiré superlattice vector $\bm{G}$. These are superpositions of the reciprocal superlattice basis vectors $\bm{b}_j^{(\rm m)}$ corresponding to the superlattice basis vectors in Eq.~\eqref{eq:moire_basis}, i.e., with $\bm{a}_{i}^{(\rm m)}\cdot \bm{b}_{j}^{(\rm m)}=2\pi \delta_{ij}$.

In this work we consider the moiré potential as an additional set of parameters. Considering the 120$^{\circ}$ rotation symmetry of the TMDC honeycomb lattice that is inherited by the moiré superlattice, as well as the fact that $V_{\rm m}$ needs to be hermitian, the only permissible form of the $V_{\bm{G}}$ is given by~\cite{wu2018theory,gotting2022moire}
\begin{equation}\label{eq:V_G}
	V_{\bm{G}}=\left\lbrace\begin{matrix}\mathcal{V}e^{i\Psi}& \bm{G}=\bm{G}_{1,3,5}\\
		\mathcal{V}e^{-i\Psi}& \bm{G}=\bm{G}_{2,4,6}\\
		0&\text{else}\end{matrix}\right.
\end{equation}
when restricting to nearest-neighbor coupling in reciprocal space. Here, the $\bm{G}_j$ point to the nearest neighbor MBZs of the first MBZ, numbered anticlockwise. The parameter $\mathcal{V}$ determines the overall strength of the moiré potential, while the phase $\Psi$ determines its shape, i.e., the position of maxima and minima. 

The total moiré exciton Hamiltonian $H_{\rm ex-hom}+V_{\rm m}$ describes \textit{homogeneous} excitons which are scattered by the moiré potential, changing their total wave vector $\bm{K}$ by a reciprocal moiré superlattice vector $\bm{G}$. By writing the total exciton wave vector as $\bm{K}=\bm{k}+\bm{G}'$ with some moiré reciprocal superlattice vector $\bm{G}'$ and $\bm{k}$ within the first MBZ, we can see that the moiré potential only leads to interactions for homogeneous excitons with the same value of $\bm{k}$ within the first MBZ, which therefore remains a good quantum number. This property, originating from the superlattice periodicity of the moiré potential, allows us to construct moiré exciton operators analogously to Bloch electrons in a periodic potential~\cite{brem2020tunable, knorr2022exciton, jurgens2024theory}
\begin{equation}\label{eq:X_dag}
	X_{n,\bm{k}}^{\dagger}=\sum_{\bm{G}}\varphi_{\bm{k}+\bm{G}}^{(n)} Y^{\dagger}_{\bm{k}+\bm{G}}\,,
\end{equation}
where the expansion coefficients fulfill the eigenvalue equation
\begin{equation}\label{eq:eigenvalue_moire}
	\sum_{\bm{G}'}\qty(E_{\bm{k}+\bm{G}'}\delta_{\bm{G},\bm{G}'}+V_{\bm{G}-\bm{G}'})\varphi_{\bm{k}+\bm{G}'}^{(n)}=\hbar\omega_{n,\bm{k}}\varphi_{\bm{k}+\bm{G}}^{(n)}\,.
\end{equation}
This equation follows from demanding the moiré exciton operators to create eigenstates of $H_{\rm ex-hom}+V_{\rm m}$, i.e., 
\begin{equation}\label{eq:H_ex}
	H_{\rm ex}\equiv H_{\rm ex-hom}+V_{\rm m}=\sum_{n,\bm{k}}\hbar\omega_{n,\bm{k}}X^{\dagger}_{n,\bm{k}}X^{}_{n,\bm{k}}
\end{equation}
in the low-densi\-ty limit analogously to Eq.~\eqref{eq:ex-hom}. Since the expansion coefficients $\varphi_{\bm{k}+\bm{G}}^{(n)}$ obey the hermitian eigenvalue equation \eqref{eq:eigenvalue_moire}, they form a complete and orthonormal set of functions for each $\bm{k}$ within the first MBZ. This also implies that the moiré exciton annihilation $X_{n,\bm{k}}$ and creation operators 	$X_{n,\bm{k}}^{\dagger}$ inherit the bosonic commutation relations in the low density limit from the \textit{homogeneous} excitons (see App.~\ref{app:moire_exciton_picture}). Note that while the expansion coefficients $\varphi_{\bm{K}}^{(n)}$ are defined for any $\bm{K}$, Eq.~\eqref{eq:eigenvalue_moire} ensures that the dispersion relation automatically satisfies $\omega_{n,\bm{k}+\bm{G}}=\omega_{n,\bm{k}}$ and Eq.~\eqref{eq:X_dag} ensures that the moire exciton operators automatically satisfy $X_{n,\bm{k}+\bm{G}}^{\dagger}=X_{n,\bm{k}}^{\dagger}$.

\begin{figure}
	\centering
	\includegraphics[width=\linewidth]{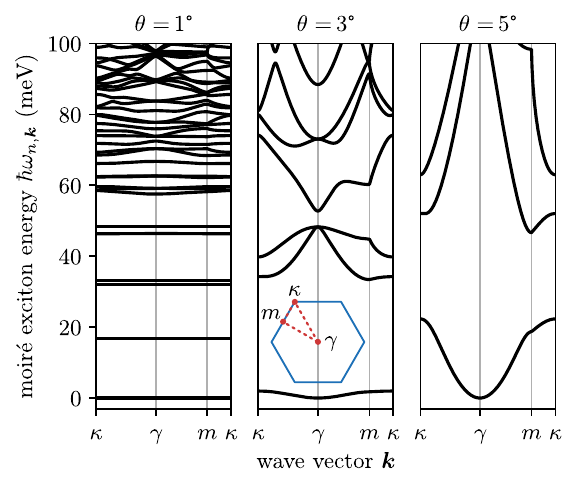}
	\caption{Moir\'e exciton band structures for different twist angles along high symmetry directions in the first MBZ as indicated in the inset.}
	\label{fig:band_structure}
\end{figure}

Since the twist angle $\theta$ changes the size of the MBZ and the length of the reciprocal superlattice vectors $\bm{G}$, also the moiré exciton bandstructure $\hbar\omega_{n\bm{k}}$, determined via Eq.~\eqref{eq:eigenvalue_moire}, is strongly impacted by the twist angle~\cite{brem2020tunable, knorr2022exciton, mak2022semiconductor, jurgens2024theory}. Figure~\ref{fig:band_structure} shows exemplary moiré exciton band structures for a small twist angle of $\theta=1^{\circ}$ (left), an intermediate twist angle of $\theta=3^{\circ}$ (center), and a large twist angle of $\theta=5^{\circ}$ (right) along high-symmetry directions in the first MBZ as shown in the inset. We denote high symmetry points in the MBZ with small letters to distinguish them from the conventional Brillouin zone, i.e., $\gamma$ corresponds to $\Gamma$, $\kappa$ to K and $m$ to M in terms of the usual terminology. The parameters that we use for the calculations are discussed in Sec.~\ref{sec:parameters}. For small twist angles (left), the moiré exciton bands are virtually flat, such that the moiré superlattice hosts localized excitons which have an essentially infinite effective mass. When increasing the twist angle, the dispersion becomes steeper until the lowest band shows a visibly quadratic dispersion around the MBZ center at large twist angles (right). In this case the moiré excitons have a finite effective mass and can move through the superlattice more freely~\cite{brem2020tunable, knorr2022exciton}. This distinction will have a huge impact on absorption spectra and coupling to pho\-nons throughout this work. Note that we chose the zero point of the energy in Fig.~\ref{fig:band_structure} such that the $\gamma$-point of the lowest lying moiré exciton band has vanishing energy $\hbar\omega_{1,\bm{k}=\bm{0}}=0$. This convention will be used throughout the paper.

\subsection{Moiré exciton-pho\-non coupling}
To thoroughly model the absorption line shape of moiré excitons, one needs to take into account the coupling with pho\-nons, which will turn out to have a large effect. The coupling of the \textit{homogeneous} excitons to pho\-nons is modeled via the standard coupling, linear in the lattice displacement~\cite{krummheuer2002theory,selig2016excitonic,lengers2020theory},
\begin{equation}\label{eq:V_ex_ph}
	V_{\rm ex-ph}=\sum_{j,\bm{K},\bm{Q}} \hbar g_{j,\bm{Q}}Y_{\bm{K}+\bm{Q}}^{\dagger}Y_{\bm{K}}^{}\qty(b_{j,\bm{Q}}^{}+b_{j,-\bm{Q}}^{\dagger})
\end{equation} 
with the coupling constants
\begin{equation}\label{eq:g_jQ}
	g_{j,\bm{Q}}=g_{j,\bm{Q}}^{(\rm e)}F(\mu_h\bm{Q})-g_{j,\bm{Q}}^{(\rm h)}F(-\mu_e\bm{Q})\,.
\end{equation}
Here, $g_{j,\bm{Q}}^{(\rm e/h)}$ are the coupling constants for pho\-nons scattering with electrons/holes and
\begin{equation}
	F(\bm{Q})=\sum_{\bm{K}}\Phi(\bm{K})\Phi^*(\bm{K}+\bm{Q})
\end{equation}
is the homogeneous exciton form factor with $\Phi(\bm{K})$ being the wavefunction of the homogeneous 1s exciton in $\bm{K}$-space (for details see App.~\ref{app:exciton_picture}). The coupling constants in Eq.~\eqref{eq:g_jQ} contain the mass fractions of electron and hole, $\mu_{\rm e/h}=m_{\rm e/h}/M$, where $m_{\rm e/h}$ is the effective mass of the electron/hole. The bosonic operators $b_{j,\bm{Q}}^{(\dagger)}$ in Eq.~\eqref{eq:V_ex_ph} destroy (create) pho\-nons in branch $j$ with momentum $\hbar\bm{Q}$ and the free dynamics of the pho\-nons is described by the Hamiltonian
\begin{equation}\label{eq:H_ph}
	H_{\rm ph}=\sum_{j,\bm{Q}}\hbar\Omega_{j,\bm{Q}}b_{j,\bm{Q}}^{\dagger}b_{j,\bm{Q}}^{}
\end{equation}
with the pho\-non dispersion relation $\Omega_{j,\bm{Q}}$.

To describe the interaction between moiré excitons and pho\-nons we invert Eq.~\eqref{eq:X_dag} using the completeness of the coefficients $\varphi_{\bm{k}+\bm{G}}^{(n)}$ to obtain~\cite{brem2020tunable, knorr2022exciton, jurgens2024theory} (for details see App.~\ref{app:moire_exciton_picture})
\begin{align}\label{eq:V_ex_ph_moire}
	V_{\rm ex-ph}&=\underset{\bm{k},\bm{Q}}{\sum_{n,n',j}} \hbar \mathcal{G}^{(n,n')}_{j,\bm{k},\bm{Q}} X_{n',\bm{k}+\bm{Q}}^{\dagger}X_{n,\bm{k}}^{}\notag\\
	&\qquad\times\qty(b_{j,\bm{Q}}^{}+b_{j,-\bm{Q}}^{\dagger})\,.
\end{align} 
Since the dependence of the moire exciton operators on $\bm{k}$ is periodic with respect to the reciprocal lattice, $\bm{k}+\bm{Q}$ in $X_{n',\bm{k}+\bm{Q}}^{\dagger}$ is equivalent to $\bm{k}+\bm{Q}-\bm{G}$ for some reciprocal superlattice vector $\bm{G}$ such that it lies in the first MBZ. The interaction with pho\-nons thus potentially leads to Umklapp processes for the moiré excitons, which becomes increasingly important with decreasing twist angle because of the decreasing size of the MBZ [see Eq.~\eqref{eq:moire_basis}]. The coupling constants for the moiré exciton-pho\-non coupling are given by
\begin{equation}\label{eq:g_moire}
	\mathcal{G}^{(n,n')}_{j,\bm{k},\bm{Q}}=g_{j,\bm{Q}}f_{\bm{k},\bm{Q}}^{(n,n')}
\end{equation}
with the moiré exciton form factors
\begin{equation}\label{eq:formfactor_moire}
	f_{\bm{k},\bm{Q}}^{(n,n')}=\sum_{\bm{G}}\qty(\varphi^{(n')}_{\bm{k}+\bm{Q}+\bm{G}})^*\varphi_{\bm{k}+\bm{G}}^{(n)}\,.
\end{equation}
Note that these are, in analogy to $\omega_{n,\bm{k}}$ and $X^{(\dagger)}_{n,\bm{k}}$, by definition reciprocal superlattice-periodic with respect to $\bm{k}$, fulfilling $f_{\bm{k}+\bm{G},\bm{Q}}^{(n,n')}=f_{\bm{k},\bm{Q}}^{(n,n')}$.
\subsection{Optical driving and dipole matrix elements of moiré excitons}
To be able to model absorption spectra, we need to include optical driving of the moiré excitons via an external classical light source. Considering illumination perpendicular to the sample surface, such that only homogeneous excitons with vanishing momentum $\hbar\bm{K}=\bm{0}$ are created, the interaction between a laser and the homogeneous excitons is given by~\cite{krummheuer2002theory,lengers2020theory}
\begin{equation}\label{eq:V_ex_laser_original}
	V_{\rm ex-laser}(t)=-\mathcal{E}(t)\tilde{M}Y_{\bm{0}}^{\dagger}+h.c.
\end{equation}
in dipole and rotating-wave approximation. Here $\tilde{M}$ is the dipole matrix element of the homogeneous exciton projected onto the polarization of the laser and  $\mathcal{E}(t)$ is the positive frequency component of the laser field. We assume that the laser's polarization and spectrum are chosen such that only homogeneous intra\-layer excitons at the K valley of one of the monolayers are excited~\cite{xiao2012coupled, wang2018colloquium, ovesen2019interlayer}. Inverting Eq.~\eqref{eq:X_dag}, analogously to the derivation of Eq.~\eqref{eq:V_ex_ph_moire} we obtain
\begin{equation}\label{eq:V_ex_laser}
	V_{\rm ex-laser}(t)=-\sum_n \mathcal{E}(t) M_n X_{n,\bm{0}}^{\dagger}+h.c.
\end{equation}
for the coupling between an external laser impinging perpendicular to the sample surface and intra\-layer moiré excitons. Here, 
\begin{equation}\label{eq:def_M_n}
	M_n=\qty(\varphi_{\bm{0}}^{(n)})^*\tilde{M}
\end{equation}
is the dipole matrix element for the excitation of moiré excitons in the $n$-th band and the laser only creates moiré excitons with vanishing momentum $\hbar\bm{k}=\bm{0}$ due to the perpendicular illumination. Note that we can discard the details of the dipole moment of the homogeneous exciton $\tilde{M}$ in the following since we are interested only in the linear optical regime, where it leads to a trivial scaling of all spectra.

\subsection{Modeling absorption spectra}
The complex positive frequency component of the macroscopic polarization of the moiré exciton system can be deduced from the interaction in Eq.~\eqref{eq:V_ex_laser} as~\cite{koch2001correlation, haug2009quantum, groll2025fundamentals}
\begin{equation}\label{eq:P_makro}
	P(t)=\frac{1}{V}\sum_n M_n^* \expval{X_{n,\bm{0}}}(t)
\end{equation}
with $V$ denoting a normalization volume. In linear response theory the macroscopic polarization and the excitation via the laser are connected as~\cite{griffiths2005introduction, klingshirn2012semiconductor}
\begin{equation}
	\tilde{P}(\omega)=\epsilon_0 \chi(\omega)\tilde{\mathcal{E}}(\omega)
\end{equation}
with the vacuum permittivity $\epsilon_0$, the linear susceptibility $\chi(\omega)$ and the Fourier transforms of the laser field $\tilde{\mathcal{E}}$ and macroscopic polarization $\tilde{P}$. The linear absorption spectrum $\alpha(\omega)$ is then given by~\cite{mahan2000many, krummheuer2002theory}
\begin{equation}\label{eq:alpha_omega_1}
	\alpha(\omega)\sim \Im\qty[\chi(\omega)]\sim \Im\qty[\tilde{P}(\omega)/\tilde{\mathcal{E}}(\omega)]\,.
\end{equation}
To obtain the linear absorption spectrum we therefore need to determine the dynamics of the microscopic moiré exciton polarizations $\expval{X_{n,\bm{k}}}(t)$. To calculate the macroscopic polarization in Eq.~\eqref{eq:P_makro}, only $\bm{k}=\bm{0}$ is relevant, which is also the only component that is driven by the laser [see Eq.~\eqref{eq:V_ex_laser}]. Considering a thermal pho\-non bath, weak optical driving and a second order Born as well as a TCL approximation, closed equations of motion for the microscopic moiré exciton polarizations, called the TCL master equation~\cite{breuer2002theory,lengers2020theory, jurgens2024theory}, are obtained. As shown in the detailed derivation in App.~\ref{app:TCL}, the equations of motion for the microscopic polarizations $\expval{X_{n,\bm{k}}}(t)$ with different $\bm{k}$ decouple, such that in the following we can focus on the optically active microscopic polarizations $p_n(t)=\expval{X_{n,\bm{0}}}(t)$ obeying the TCL master equation
\begin{align}
	\dv{t}p_{n}(t)&=-i\omega_{n}p_{n}(t)+\frac{i}{\hbar}\mathcal{E}(t)M_n\notag\\
	&-\sum_{n'',j}\Gamma_{j}^{(n,n'')}(t)p_{n''}(t)-\frac{\gamma_n}{2}p_{n}(t)\,.\label{eq:TCL}
\end{align}
Radiative decay of the moiré excitons is included phenomenologically with the rates
\begin{equation}\label{eq:gamma_n}
	\gamma_n=|\varphi_{\bm{0}}^{(n)}|^2\gamma_{\rm h}\,,
\end{equation}
where $\gamma_{\rm h}$ is the decay rate of the homogeneous exciton occupation. These decay rates account for the correct dependence on the dipole matrix elements $M_n$, however we neglect any dependence on the moiré exciton transition frequency $\omega_{n}=\omega_{n,\bm{0}}$ assuming a sufficiently flat density of states for the emitted photons in the relevant range of the $\omega_{n}$~\cite{weisskopf1930berechnung,novotny2012principles}. This is justified since we will be interested in a range of moiré exciton energies on the order of 100~meV around the energy of the homogeneous 1s exciton which is on the order of 1.5~eV~\cite{ovesen2019interlayer, wietek2024nonlinear}. Note that the decay rates fulfill $\sum_n \gamma_n=\gamma_{\rm h}$ due to the completeness of the coefficients $\varphi_{\bm{k}+\bm{G}}^{(n)}$ (see App.~\ref{app:moire_exciton_picture}). This implies that in our phenomenological decay mo\-del, the decay rate of the bright homogeneous exciton is simply redistributed to the different bright moiré excitons at the $\gamma$-point.

All effects due to the thermal pho\-non bath, i.e., pho\-non-in\-duced dissipation and energy renormalization (polaron shifts), are captured in the time dependent dissipation coefficient matrix 
\begin{equation}\label{eq:Gamma_t}
	\Gamma_{j}^{(n,n'')}(t)=\int\dd\Omega\, \rho_{j}^{(n,n'')}(\Omega)\int\limits_0^t\dd\tau\, e^{-i\Omega\tau}
\end{equation}
due to pho\-nons in branch $j$. The different pho\-non branches enter additively in Eq.~\eqref{eq:TCL} due to the second order Born approximation and the assumption of a thermal pho\-non state applied in its derivation. The impact of the pho\-nons is completely captured in the generalized pho\-non spectral density (gPSD)~\cite{jurgens2024theory}
\begin{align}\label{eq:g_PSD}
	\rho_{j}^{(n,n'')}(\Omega)&=\sum_{n',\bm{Q},\sigma=\pm}\qty(\mathcal{G}_{j,\bm{0},-\bm{Q}}^{(n,n')})^*\mathcal{G}_{j,\bm{0},\bm{-Q}}^{(n'',n')}N_{j,\bm{Q}}^{(-\sigma)}\times\\
	&\quad\times\delta\qty(\Omega+\sigma\Omega_{j,-\sigma\bm{Q}}+\omega_{n'',\bm{0}}-\omega_{n',-\bm{Q}})\notag
\end{align}
with
\begin{equation}\label{eq:N_phon}
	N_{j,\bm{Q}}^{(\mp)}=\frac{1}{2}\mp \frac{1}{2}+n_{j,\mp\bm{Q}}
\end{equation}
describing the impact of temperature $T$ on absorption/emission of pho\-nons via the thermal pho\-non distribution
\begin{equation}\label{eq:n_phon}
	n_{j,\bm{Q}}=\frac{1}{\exp(\frac{\hbar\Omega_{j,\bm{Q}}}{k_{\rm B}T})-1}\,.
\end{equation}
The gPSD in Eq.~\eqref{eq:g_PSD} describes pho\-non-in\-duced transitions between moiré exciton bands in second order of the pho\-non coupling. Its value at $\Omega=0$ determines the strength of energy conserving transitions between the involved moiré exciton states with energies $\omega_{n'',\bm{0}}$ and $\omega_{n',-\bm{Q}}$, as described by the $\delta$-function, while all contributions with $\Omega\neq 0$ are due to pho\-non-in\-duced transitions that do not conserve energy in the exciton-pho\-non system. For this reason we refer to $\hbar\Omega$ simply as the energy mismatch.

The TCL master equation~\eqref{eq:TCL} is local in time, i.e., its right hand side involves only microscopic polarizations $p_n(t)$ with the same time argument. However, it is still a non-Markovian equation of motion since the dissipation coefficients in Eq.~\eqref{eq:Gamma_t} contain memory effects via the explicit dependence on the initial preparation by optical excitation at $t=0$. This memory effect leads to the time dependence of these dissipation coefficients, which however become time-independent if we take the long-time limit $t\rightarrow \infty$. This constitutes the Markov limit of the equations of motion where memory effects vanish~\cite{breuer2002theory}.

Transitions with a non-vanishing energy mismatch $\Omega\neq 0$ in Eqs.~\eqref{eq:Gamma_t} and \eqref{eq:g_PSD} impact the moiré exciton polarization dynamics on short time scales but in general lose impact in the long time, i.e., Markov, limit. In this limit the pho\-non-in\-duced dissipation coefficients are given by~\cite{breuer2002theory, jurgens2024theory}
\begin{align}\label{eq:markov}
	\overline{\Gamma}_{j}^{(n,n'')}&=	\lim\limits_{t\rightarrow\infty}\Gamma_{j}^{(n,n'')}(t)\\
	&=-i\mathcal{P}\int\dd\Omega\, \frac{\rho_{j}^{(n,n'')}(\Omega)}{\Omega}+\pi \rho_{j}^{(n,n'')}(0)\notag\,,
\end{align}
where $\mathcal{P}$ denotes the principal value of the integral and we made use of the Dirac identity. Considering the TCL master equation~\eqref{eq:TCL}, it is clear that the pho\-non-in\-duced damping of polarizations in the Markov limit is described by ${\rm Re}\big(\overline{\Gamma}_{j}^{(n,n'')}\big)$ and therefore determined solely by the gPSD at $\Omega=0$, i.e., by energy conserving pho\-non-in\-duced transitions between the moiré exciton bands. ${\rm Im}\big(\overline{\Gamma}_{j}^{(n,n'')}\big)$ on the other hand, which describes energy renormalizations, i.e., polaron shifts~\cite{krummheuer2002theory,lengers2020theory, preuss2022resonant}, is impacted by the full gPSD and vanishes if it is symmetric in $\Omega$.

Note that the diagonal part of the dissipation coefficient matrix in $\Gamma_{j}^{(n,n'')}$ in the TCL master equation~\eqref{eq:TCL} describes time dependent energy renormalizations via its imaginary part and time dependent damping rates via its real part. This simple interpretation is however only valid if the off-diagonal elements are small. Otherwise the coupling between different bands in Eq.~\eqref{eq:TCL}, called inter-polarization coupling in the following, potentially leads to hybridization of moiré exciton bands induced by the coupling to pho\-nons. Still, to better understand the role of the dissipation coefficients, it is instructive to consider the diagonal part of the Markov limit decay rates
\begin{align}
	\text{Re}\qty[\overline{\Gamma}_{j}^{(n,n)}]&=\pi \rho_{j}^{(n,n)}(0)\notag\\
	&=\pi\sum_{n',\bm{Q},\sigma=\pm}\qty|\mathcal{G}_{j,\bm{0},\bm{-Q}}^{(n,n')}|^2N_{j,\bm{Q}}^{-\sigma}\times\notag\\
	&\quad\times\delta\qty(\sigma\Omega_{j,-\sigma\bm{Q}}+\omega_{n,\bm{0}}-\omega_{n',-\bm{Q}})\,.
\end{align}
This decay rate corresponds to one half of Fermi's golden rule transition rate calculated via the potential in Eq.~\eqref{eq:V_ex_ph_moire} for an initial moiré exciton in the $n$-th band with momentum $\bm{0}$. The time dependent dissipation coefficients in Eq.~\eqref{eq:TCL} extend this result to the general non-Markov\-ian case. Note that the factor of one half is due to the fact that the exciton occupation in the low density limit is given by the absolute square of the interband polarization. 

In the following we will use the TCL master equation~\eqref{eq:TCL} to determine linear absorption spectra via Eq.~\eqref{eq:alpha_omega_1}. This is particularly easy when considering a weak ultrashort excitation $\mathcal{E}(t)=\mathcal{E}_0\delta(t)$ with $\mathcal{E}_0\in\mathbb{R}$~\cite{vagov2002electron, krummheuer2002theory}. This implies a broad excitation spectrum $\tilde{\mathcal{E}}(\omega)=\mathcal{E}_0$, such that $\alpha(\omega)\sim \Im[\tilde{P}(\omega)]$. Assuming vanishing moiré exciton polarizations before the ultrashort excitation, we obtain the polarizations directly after via integrating Eq.~\eqref{eq:TCL} from $t=0^-$ to $t=0^+$, leading to
\begin{equation}\label{eq:initia_p}
	p_{n}(t=0^+)=\frac{i}{\hbar}\mathcal{E}_0M_n\,.
\end{equation}
We will treat this as the initial condition for the moiré exciton polarization dynamics, which we then determine via the TCL master equation~\eqref{eq:TCL} for vanishing optical driving $\mathcal{E}(t)=0$ for $t>0$. Note that we already implicitly assumed this in the derivation of the time dependent coefficient matrix in Eq.~\eqref{eq:Gamma_t}, as can be seen by the lower limit of the $\tau$-integral starting at $\tau=0$ (for details, see App.~\ref{app:TCL}).

With these considerations we finally arrive at the linear absorption spectrum 
\begin{equation}\label{eq:alpha_omega}
	\alpha(\omega)\sim \sum_n \Im\qty[\int\limits_0^{\infty}\dd t\, e^{i\omega t} M_n^* p_{n}(t)]
\end{equation}
with the initial condition for the microscopic polarizations in Eq.~\eqref{eq:initia_p} and their dynamics determined via the TCL master equation~\eqref{eq:TCL} for $\mathcal{E}(t>0)=0$.
\subsection{Parameters for modeling MoSe$_2$ intra\-layer excitons in a twisted MoSe$_2$/WSe$_2$ heterobilayer}\label{sec:parameters}
To apply our theoretical model exemplarily to a representative material, we consider parameters that are typical for a twisted MoSe$_2$/WSe$_2$ heterobilayer. This material hosts intra\-layer excitons in both monolayers, as well as inter\-layer excitons~\cite{ovesen2019interlayer,wietek2024nonlinear}. The latter constitute the energetically lowest-lying excitons and therefore dominate photoluminescence spectra but have a negligible oscillator strength in our context due to their out-of plane dipole moment, such that they do not contribute strongly to absorption spectra. We will focus here on the absorption spectra of the 1s intra\-layer exciton in MoSe$_2$ and how it is affected by the presence of the moiré superlattice. This intra\-layer exciton lies roughly 100~meV below the 1s intra\-layer exciton in WSe$_2$ and more than 100~meV above the 1s inter\-layer exciton. Assuming sufficiently resonant excitation of the 1s intra\-layer exciton in MoSe$_2$ allows us to reduce the excitonic landscape and apply the model developed in the previous sections.

This discussion implies that the absorption spectra that we present in the following make only useful experimental predictions in a range of roughly $\pm 100$~meV relative to the lowest lying exciton in our model. This has to be kept in mind when comparing these simulations to actual data. In this sense we present simulations on the effect of pho\-non scattering and the moiré superlattice on the MoSe$_2$ intra\-layer exciton absorption peak in a twisted MoSe$_2$/ WSe$_2$ heterobilayer.

Note that intra\-layer to inter\-layer exciton conversion is not considered in our model. This process impacts incoherent excitons on longer time scales while the absorption spectrum is determined completely by the dynamics of coherent excitons on shorter time scales~\cite{ovesen2019interlayer}. However it could lead to additional dephasing and broadening of all calculated spectra presented in this work.

For the electrons and holes in MoSe$_2$ we consider the effective masses $m_{\mathrm{e}}=0.49 m_0$ and $m_{\mathrm{h}}=0.61m_0$ with $m_0$ being the free electron mass~\cite{rasmussen2015computational,lengers2020theory}. These electrons and holes are coupled to two effective pho\-non branches in the MoSe$_2$ monolayer~\cite{li2013intrinsic,jin2014intrinsic, lengers2020theory}, assumed to be unaffected by the moiré potential. We consider an acoustic pho\-non branch with the linear dispersion relation
\begin{align}\label{eq:ac_disp}
	\Omega_{j=\mathrm{ac},\bm{Q}}=c_s|\bm{Q}|
\end{align}
and an effective dispersion-less optical pho\-non branch
\begin{align}\label{eq:opt_disp}
	\Omega_{j=\mathrm{opt},\bm{Q}}=\Omega_{\mathrm{opt}}\,,
\end{align}
both coupled to the exciton in MoSe$_2$ via the deformation potential coupling
\begin{align}
	g_{j,\bm{Q}}^{(\rm{e/h})}=\sqrt{\frac{1}{2A\hbar\rho\Omega_{j,\bm{Q}}}}\Delta V_{j,\bm{Q}}^{(\rm{e/h})}\,
\end{align}
with all relevant coupling parameters from Ref.~\cite{lengers2020theory} based mostly on Ref.~\cite{jin2014intrinsic}, as discussed therein. Here, $A$ is the two-dimensional normalization volume and $\rho=4.26\times 10^{-7}$~g$/$cm$^2$ is the mass density of MoSe$_2$. The sound velocity of the acoustic branch in MoSe$_2$ is given by $c_s=4.1$~nm$/$ps and the optical pho\-non energy is $\hbar\Omega_{\mathrm{opt}}=34.4$~meV. The deformation potential for the acoustic pho\-nons is modeled via
\begin{align}
	\Delta V_{\mathrm{ac},\bm{Q}}^{(\rm{e/h})}=D_{\mathrm{ac}}^{(\rm{e/h})}|\bm{Q}|
\end{align}
with deformation potential constants $D_{\mathrm{ac}}^{(\rm{e})}=2.40$~eV and $D_{\mathrm{ac}}^{(\rm{h})}=-1.98$~eV for electrons and holes, respectively. The deformation potential for optical pho\-nons is modeled via
\begin{align}
	\Delta V_{\mathrm{opt},\bm{Q}}^{(\rm{e/h})}=D_{\mathrm{opt}}^{(\rm{e/h})}
\end{align}
with deformation potential constants $D_{\mathrm{opt}}^{(\rm{e})}=52\frac{\text{eV}}{\text{nm}}$ and $D_{\mathrm{opt}}^{(\rm{h})}=-49\frac{\text{eV}}{\text{nm}}$ for electrons and holes, respectively. For a discussion on the signs of the deformation potential constants, see Ref.~\cite{lengers2020theory}. The form factor of the homogeneous 1s exciton, which is needed to determine the exciton-pho\-non coupling, is approximated as a Gaussian
\begin{align}\label{eq:exc_FF}
	F(\bm{Q})=e^{-\frac{1}{2}|\bm{Q}|^2\sigma^2}
\end{align}
whose width of $\sigma=1$~nm corresponds to typical 1s MoSe$_2$ intra\-layer exciton wavefunction extensions~\cite{ovesen2019interlayer}. 

To determine the properties of the moiré excitons, we consider a lattice constant for both monolayers of $|\bm{a}_j|=0.332$~nm~\cite{rasmussen2015computational} and the moiré potential parameters $\mathcal{V}=11.8$~meV and $\Psi=79.5^{\circ}$~\cite{wu2018theory,gotting2022moire}. Note that there is a large variation in the intra\-layer potential parameters given in the literature, depending on the material combination and the specific stacking~\cite{yu2017moire,zhang2018moire,lin2023remarkably}. With our set of parameters we obtain intra\-layer moiré exciton band structures that agree qualitatively with those in Ref.~\cite{brem2020tunable}. We do not expect large qualitative differences in the results when varying the potential parameter $\mathcal{V}$ on the order of $\sim 1$~meV.

Finally, the radiative decay rate of the homogeneous exciton is chosen as $\gamma_{\mathrm{h}}=0.25$~ps$^{-1}$~\cite{wang2014valley, lagarde2014carrier}.
\section{Absorption properties and polarization dynamics of the lowest-lying moiré exciton band}
In this section we will focus on the polarization dynamics and absorption spectra of the lowest-lying moiré exciton band, labeled as $n=1$ in the following. This study will help us in understanding the separate impact of acoustic pho\-nons and optical pho\-nons while keeping the excitonic structure as simple as possible. To this aim, we solve the TCL master equation~\eqref{eq:TCL} for $n=1$, discarding inter-polarization coupling, i.e., the sum over $n''\neq 1$. The only relevant dissipation coefficients are therefore $\Gamma_{j}^{(1,1)}(t)$ and the only relevant gPSDs are 
\begin{align}\label{eq:g_PSD_1}
	\rho_{j}^{(1,1)}(\Omega)&=\sum_{n',\bm{Q},\sigma=\pm}\qty|\mathcal{G}_{j,\bm{0},\bm{-Q}}^{(1,n')}|^2N_{j,\bm{Q}}^{-\sigma}\\
	&\quad\times\delta\qty(\Omega+\sigma\Omega_{j,-\sigma\bm{Q}}+\omega_{1,\bm{0}}-\omega_{n',-\bm{Q}})\notag
\end{align}
describing the impact of pho\-non-in\-duced transitions from the first band to any other moiré exciton band for a given pho\-non branch $j$.

As described in App.~\ref{app:markov_non_markov}, we can separate the relevant polarization dynamics $p_{1}(t)$ in Eq.~\eqref{eq:TCL} into a Markov\-ian and a non-Markov\-ian contribution. We obtain after some analysis that, given the restrictions discussed at the beginning of this section, the dynamics after optical excitation, i.e., for $\mathcal{E}(t>0)=0$, are given by
\begin{subequations}\label{eq:pol_1}
	\begin{align}
		p_1(t)&=e^{-i\tilde{\omega}_{1}t-\frac{\tilde{\gamma}_1}{2} t}e^{-\phi_1(t)}p_{1}(0)\,,\\
		\tilde{\omega}_1&=\omega_{1}+\sum_j \Im\qty(\overline{\Gamma}^{(1,1)}_{j})\,,\\
		\tilde{\gamma}_1&=\gamma_1+2\sum_j \Re\qty(\overline{\Gamma}^{(1,1)}_{j})\,,\\
		\phi_1(t)&=\lim\limits_{\eta\rightarrow 0^+}\int\dd\Omega\, \sum_j\frac{\rho_{j}^{(1,1)}(\Omega)}{\qty(\Omega-i\eta)^2}\notag\\
		&\qquad\times\qty[1-e^{-i(\Omega-i\eta) t}]\,,\label{eq:phi_1}
	\end{align}
\end{subequations}
where the limit $\eta\rightarrow 0^+$ is taken at the end of calculations. This has the same structure as the polarization dynamics of a single two-level emitter coupled to an arbitrary number of pho\-non modes described within the independent boson model~\cite{mahan2000many,nazir2016modelling,wigger2019phonon, preuss2022resonant}. For a single completely flat band this analogy becomes exact, since the gPSD then reads 
\begin{align}\label{eq:g_PSD_flat}
	\rho_{j}^{(1,1)}(\Omega)&=\sum_{\bm{Q},\sigma=\pm}\qty|\mathcal{G}_{j,\bm{0},\bm{-Q}}^{(1,1)}|^2N_{j,\bm{Q}}^{-\sigma}\notag\\
	&\qquad\times\delta\qty(\Omega+\sigma\Omega_{j,-\sigma\bm{Q}})\,,
\end{align}
which is simply the common pho\-non spectral density of the independent boson model, however already modified by including the thermal pho\-non occupation. The latter is usually left out of the definition of the spectral density in the context of the independent boson model and appears then explicitly in the non-Markov\-ian dephasing function $\phi_1(t)$ from Eq.~\eqref{eq:phi_1}~\cite{nazir2016modelling,wigger2019phonon, preuss2022resonant}.

The structural similarity between the polarization dynamics in Eqs.~\eqref{eq:pol_1} and the one encountered within the independent boson model, separated into Markov\-ian dynamics with the frequency $\tilde{\omega}_1$ and decay rate $\tilde{\gamma}_1$ and non-Markov\-ian dynamics via the dephasing function $\phi_1(t)$, lets us perform a classification of the corresponding absorption spectra analogously to the case of solid state single photon emitters~\cite{mahan2000many,krummheuer2002theory,wigger2019phonon,preuss2022resonant}. Inserting Eqs.~\eqref{eq:pol_1} together with the initial condition given in Eq.~\eqref{eq:initia_p} into the absorption spectrum in Eq.~\eqref{eq:alpha_omega} gives rise to a power series $\alpha(\omega)\sim\sum_q \alpha^{(q)}(\omega)$ in terms of the dephasing function $\phi_1(t)$ (for details, see App.~\ref{app:zpl_psbs}).

The zeroth-order contribution with respect to the non-Markov\-ian dephasing $\phi_1(t)$ is given by
\begin{equation}
	\alpha^{(0)}(\omega)= |M_1|^2\frac{\tilde{\gamma}_1/2}{\qty(\tilde{\gamma}_1/2)^2+(\omega-\tilde{\omega}_{1})^2}\,.\label{eq:ZPL_1}
\end{equation}
This is a single Lorentzian at the polaron-shifted frequency $\tilde{\omega}_{1}$ with a width determined by the total polarization decay rate~$\tilde{\gamma}_1$, including both radiative and Markov\-ian pho\-non-in\-duced decay. This is commonly called the zero-pho\-non line (ZPL) in the context of solid state single pho\-non emitters.

The first-order contribution reads
\begin{align}
	&\alpha^{(1)}(\omega)=\mathcal{P}\!\int\!\dd\Omega \sum_j\frac{\rho_{j}^{(1,1)}(\Omega)}{\Omega^2}\notag\\
	&\qquad\qquad\qquad\qquad\times\qty[\alpha^{(0)}(\omega-\Omega)-\alpha^{(0)}(\omega)]\notag\\
	&\quad+\pi  \sum_j\frac{\rho_{j}^{(1,1)}(0)}{\tilde{\gamma}_1/2}\qty[1+(\omega-\tilde{\omega}_1)\pdv{\omega}]\alpha^{(0)}(\omega).\label{eq:PSB_1}
\end{align}
It yields pho\-non sidebands (PSBs) at $\omega=\Omega+\tilde{\omega}_1$ via the term $\alpha^{(0)}(\omega-\Omega)$, as well as modifications of the ZPL. The weight of the PSBs  is determined by the gPSD $\rho_{j}^{(1,1)}$ at the corresponding frequency mismatch $\Omega$. In the same fashion, $\alpha^{(q>1)}(\omega)$ contains higher order PSBs due to multi-pho\-non processes at $\omega=\tilde{\omega}_1+\sum_{i=1}^q \Omega_i$ with the $\Omega_i$ being the possible frequency mismatches determined by the gPSD.

With this transparent classification of absorption spectra into ZPL and PSBs at hand, in the following we will investigate the relationship between gPSD and absorption spectra of moiré excitons in detail.

\subsection{Influence of acoustic pho\-nons}

\subsubsection{gPSD for acoustic pho\-non scattering}
\begin{figure}
	\centering
	\includegraphics[width=\linewidth]{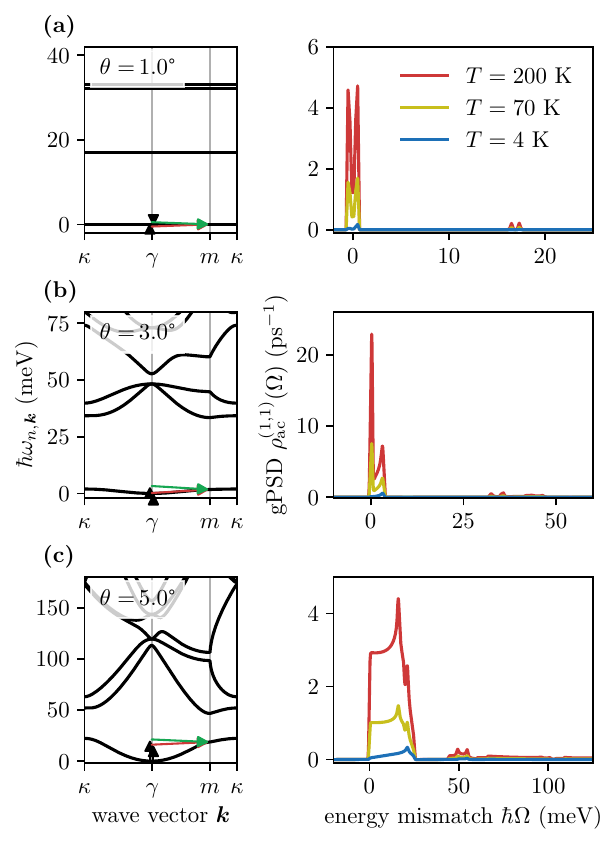}
	\caption{Moiré exciton band structure $\hbar\omega_{n,\bm{k}}$ (left) and corresponding gPSD $\rho_{j=\rm{ac}}^{(1,1)}(\Omega)$ of the lowest lying moiré exciton band due to acoustic pho\-non scattering (right) for the three twist angles $\theta=1^{\circ}$ (a), $\theta=3^{\circ}$ (b), $\theta=5^{\circ}$ (c) and three different temperatures $T=4$~K (blue), $T=70$~K (yellow), and $T=200$~K (red). Acoustic pho\-non-assisted intra\-band transitions from the $\gamma$- to the $m$-point of the lowest lying moiré exciton band (left) are sketched for phonon absorption (red arrows) and emission (green arrows) with black vertical arrows corresponding to the energy mismatch $\hbar\Omega$ that is needed for that transition to occur.}
	\label{fig:gpsddisp_ac}
\end{figure}
We begin by focusing on the impact of acoustic pho\-nons. To this aim, in Fig.~\ref{fig:gpsddisp_ac} we investigate the gPSD $\rho_{j=\rm{ac}}^{(1,1)}(\Omega)$ of the lowest lying moiré exciton band due to acoustic pho\-non scattering as a function of the energy mismatch $\hbar\Omega$. From the top to bottom row, we consider the three twist angles $\theta=1^{\circ}$ (a), $\theta=3^{\circ}$ (b), and $\theta=5^{\circ}$ (c). The left column shows the respective moiré exciton band structure, already presented in Fig.~\ref{fig:band_structure}. In addition, acoustic pho\-non-assisted intra\-band transitions of the lowest lying moiré exciton band are sketched in the following way: The absorption (emission) of an acoustic pho\-non is drawn with the quantitatively correct slope of the acoustic pho\-non dispersion relation [Eq.~\eqref{eq:ac_disp}] as a red upward (green downward) arrow. The black vertical arrows represent the energy mismatch $\hbar\Omega$ necessary for that transition to occur. Processes with a vanishing energy mismatch $\hbar\Omega=0$ correspond to energy-conserving transitions, as described by the $\delta$-function in Eq.~\eqref{eq:g_PSD_1}. We specifically show the pho\-non-in\-duced transitions from the $\gamma$-point (the optically active moiré exciton) to the $m$-point, which is a saddle point and thus corresponds to a van Hove singularity of the lowest lying moiré exciton band. Such transitions dominate the gPSDs in the right column of Fig.~\ref{fig:gpsddisp_ac} due to the correspondingly high density of states of the moiré excitons at the $m$-point. For the presented gPSDs we consider three different temperatures $T=4$~K (blue), $T=70$~K (yellow), and $T=200$~K (red).

Starting with a small twist angle of $\theta=1^{\circ}$ in Fig.~\ref{fig:gpsddisp_ac} (a), we have a particularly flat moiré exciton dispersion relation. Due to the flat band and the finite slope of the acoustic pho\-nons, the pho\-non emission transition (green arrow) from the $\gamma$- to the $m$-point requires a positive energy mismatch $\hbar\Omega>0$, while the pho\-non absorption transition (red arrow) requires a negative energy mismatch $\hbar\Omega<0$. In the gPSD on the right this leads to a characteristic double peak structure, centered around $\hbar\Omega=0$, for elevated temperatures (red, yellow). At such high temperatures, pho\-non absorption and stimulated emission dominate over spontaneous emission for the relevant acoustic modes, leading to an approximately symmetric gPSD around $\hbar\Omega=0$. For sufficiently low temperatures, here $T=4$~K (blue), the relevant pho\-non absorption and stimulated emission processes are suppressed. This leads to a suppression of the pho\-non absorption peak in the gPSD at $\hbar\Omega<0$, while the pho\-non emission peak at $\hbar\Omega>0$ is still well visible due to the possibility of spontaneous pho\-non emission. Furthermore the gPSD decreases overall with decreasing temperature due to the less efficient pho\-non scattering [thermal occupation factor in Eq.~\eqref{eq:g_PSD_1}].

In addition we find a much smaller double peak structure at $\hbar\Omega\approx 17$~meV, stemming from scattering processes involving the higher lying moiré exciton bands around $\hbar\omega_{n\bm{k}}\approx 17$~meV in Fig.~\ref{fig:gpsddisp_ac} (a). This shows that the dynamics of the moiré exciton polarization of a single band in Eq.~\eqref{eq:TCL} [here $p_1(t)$] as well as the corresponding absorption spectrum already contain an impact from the scattering into all other bands via the sum over $n'$ in the gPSD in Eq.~\eqref{eq:g_PSD}, even when neglecting inter-polarization coupling with $n''\neq n$ in the TCL master equation~\eqref{eq:TCL}. However, we found that typically pho\-non-in\-duced intra\-band scattering, i.e., $n'=n''=n$ in Eq.~\eqref{eq:g_PSD}, dominates pho\-non-in\-duced inter\-band scattering with $n'\neq n$ or $n''\neq n$. A consequence of this is, that the double peak structure around $\hbar\Omega\approx 17$~meV in Fig.~\ref{fig:gpsddisp_ac} (a) is much smaller than the one around $\hbar\Omega=0$. In the following discussion we will focus on intra\-band scattering.

Increasing the twist angle to $\theta=3^{\circ}$ in Fig.~\ref{fig:gpsddisp_ac} (b), the moiré exciton bands become curved with the lowest lying one obtaining an upwards curvature, such that both pho\-non absorption and emission processes, inducing scattering between $\gamma$- and $m$-point, require a positive energy mismatch $\hbar\Omega\gtrsim0$. Due to this, the two peaks that were centered around $\hbar\Omega=0$ in the case of flat bands in (a), are now centered around $\hbar\Omega=\hbar(\omega_{1,\bm{k}=m}-\omega_{1,\bm{k}=\gamma})>0$, i.e., they move towards larger energy mismatches $\hbar\Omega\gtrsim0$ in the gPSD. Analogous to the case of $\theta=1^{\circ}$, the gPSD increases overall with increasing temperature and the pho\-non absorption peak from the transition to the $m$-point, i.e., the one for smaller energy mismatch $\hbar\Omega$, vanishes at sufficiently low temperatures (blue).

At $\theta=5^{\circ}$ in Fig.~\ref{fig:gpsddisp_ac} (c), the double peak structure is shifted further to higher energy mismatches $\hbar\Omega$ since the energetic distance between $\gamma$- and $m$-point $\hbar(\omega_{1,\bm{k}=m}-\omega_{1,\bm{k}=\gamma})$ grows with increasing twist angle. This reveals that the gPSD around $\hbar\Omega=0$ exhibits a sudden rise reaching a plateau for elevated temperatures (red, yellow) and a slower rise for low temperatures (blue). This rise around $\hbar\Omega=0$ stems from pho\-non-assisted transitions from the $\gamma$-point to the vicinity of the $\gamma$-point, which are thus typically the relevant transitions for the Markov-limit decay rate in Eq.~\eqref{eq:markov}.

For the large twist angle of $\theta=5^{\circ}$, this behavior of the gPSD is not hidden by the impact of the van Hove singularities anymore. Similarly we can now see that the intra\-band contribution to the gPSD has a clear cutoff at $\hbar\Omega\approx 25$~meV, corresponding to the bandwidth $\hbar(\omega_{1,\bm{k}=\kappa}-\omega_{1,\bm{k}=\gamma})$ of the lowest lying moiré exciton band. All these features lead to a quite characteristic shape of the gPSD, which is directly connected to the density of states of the moiré excitons. This connection will be explored further in the context of optical pho\-non scattering.

Finally, note that the peak in the gPSD stemming from phonon absorption processes inducing transitions from the $\gamma$- to the $m$-point in Fig.~\ref{fig:gpsddisp_ac} lies at negative energy mismatch $\hbar\Omega<0$ for the small twist angle $\theta=1^{\circ}$ (a) and positive energy mismatch $\hbar\Omega>0$ for the large twist angle $\theta=5^{\circ}$ (c). In between these two twist angle values there is then a certain critical twist angle $\theta_c$, where the peak lies exactly at $\hbar\Omega=0$, contributing strongly to energy-conserving Markov processes [see Eq.~\eqref{eq:markov}]. This is the so-called \textit{magic angle for phonon scattering} already introduced in Ref.~\cite{jurgens2024theory}. For this twist angle acoustic phonon-assisted processes are particularly important and its value is determined here by the condition
\begin{equation}\label{eq:magic_angle}
	\omega_{1,\bm{k}=m}=\omega_{1,\bm{k}=\gamma}+\Omega_{\mathrm{ac},\bm{k}=m}\,,
\end{equation} 
ensuring energy conserving acoustic phonon-assisted transitions from the $\gamma$- to the $m$-point in Eq.~\eqref{eq:g_PSD_1}. For the considered parameters, we find $\theta_c= 2.9^{\circ}$, i.e., close to the situation depicted in Fig.~\ref{fig:gpsddisp_ac} (b) where the phonon absorption peak lies at $\hbar\Omega\gtrsim 0$.

\subsubsection{Polarization dynamics and time dependent decay rates}

\begin{figure}
	\centering
	\includegraphics[width=\linewidth]{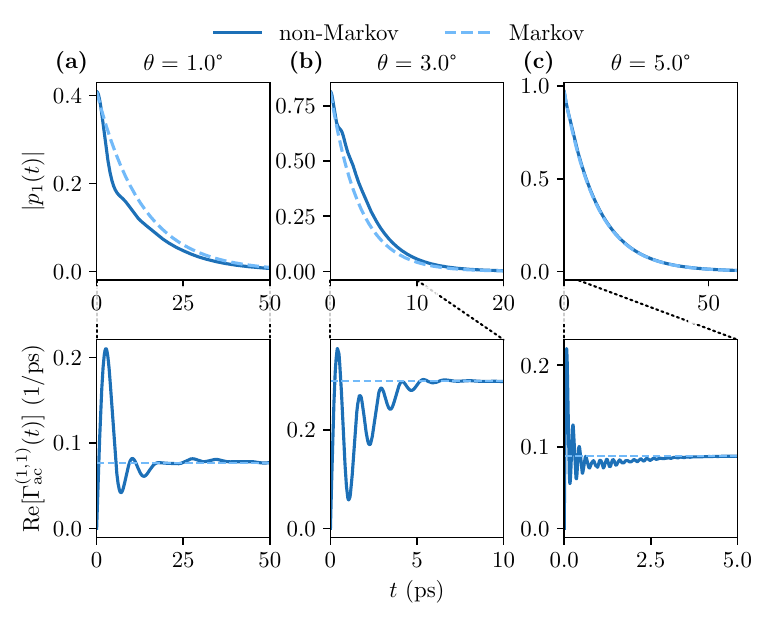}
	\caption{Dynamics of the absolute value of the moiré exciton polarization $|p_1(t)|$ from Eqs.~\eqref{eq:pol_1} including only acoustic phonon scattering $j=\rm{ac}$ (top) and corresponding time dependent decay rates $\Re\big[\Gamma_{\rm{ac}}^{(1,1)}(t)\big]$ calculated via Eq.~\eqref{eq:Gamma_t} (bottom) for the twist angles $\theta=1^{\circ}$ (a), $\theta=3^{\circ}$ (b), and $\theta=5^{\circ}$ (c) at a temperature of $T=4$~K. We compare the full simulations (solid) with the Markov limit (dashed).}
	\label{fig:dynamics_ac_4K}
\end{figure}

To investigate the impact of Markov\-ian and non-Markov\-ian pho\-non effects on the bright moiré exciton in the lowest lying band, Fig.~\ref{fig:dynamics_ac_4K} shows the dynamics of the absolute value of the moiré exciton polarization $|p_1(t)|$ from Eqs.~\eqref{eq:pol_1} (top). We include here only the impact from acoustic pho\-nons with $j=\rm{ac}$. The corresponding time dependent decay rate $\Re\big[\Gamma_{\rm{ac}}^{(1,1)}(t)\big]$, calculated via Eq.~\eqref{eq:Gamma_t}, is shown in the bottom row. In addition to the full simulations (solid), we show the results in the Markov limit (dashed), obtained by setting $\phi_1=0$ in the dynamics for $p_1$ from Eqs.~\eqref{eq:pol_1} (top) and replacing the full time dependent decay rate by $\Re\big(\overline{\Gamma}^{(1,1)}_{\rm{ac}}\big)$ (bottom), respectively. The results in Fig.~\ref{fig:dynamics_ac_4K} are calculated for $T=4$~K. Simulations for $T=70$~K and $T=200$~K can be found in App.~\ref{app:pol_dyn_T_high}. The twist angle dependent trends observed and discussed in the following mostly extend to these cases of elevated temperatures.

To evaluate the impact of the twist angle, the three columns in Fig.~\ref{fig:dynamics_ac_4K} show the results for $\theta=1^{\circ}$ (a), $\theta=3^{\circ}$ (b), and $\theta=5^{\circ}$ (c) from left to right. Starting with $\theta=1^{\circ}$ (a) we see a strong deviation between the full polarization dynamics (top, solid) and the corresponding Markov limit (top, dashed). The Markov limit clearly underestimates the decay of the polarization, as can also be seen when comparing the time dependent decay rate (bottom, solid) with the Markov limit decay rate (bottom, dashed). Furthermore, the time dependent rate shows oscillations with a period of $T\approx 8$~ps, which can be attributed to the peak in the corresponding gPSD at $\hbar\Omega=\hbar 2\pi/T\approx 0.5~$meV [Fig.~\ref{fig:gpsddisp_ac} (a)] stemming from transitions to the van Hove singularity at the $m$-point. Since these transitions dominate the gPSD at small twist angles, leading to peaks for $\hbar\Omega\neq 0$, we find strong non-Markov\-ian effects in the polarization dynamics. The fact that the Markov limit underestimates the polarization decay for $\theta=1^{\circ}$ is consistent with the correspondence between the flat band moiré system and the independent boson model, as already discussed in the context of Eq.~\eqref{eq:g_PSD_flat}. In fact the independent boson model does not produce any Markov limit decay rate for the coupling of a two-level emitter to acoustic pho\-nons via the deformation potential in two (or higher) dimensions~\cite{krummheuer2002theory, lindwall2007zero,nazir2016modelling,preuss2022resonant}.

Considering now the intermediate twist angle of $\theta=3^{\circ}$ in Fig.~\ref{fig:dynamics_ac_4K} (b), we see that the Markov limit (dashed) overestimates the polarization decay compared to the full non-Markovian simulation (solid). As discussed in the context of Eq.~\eqref{eq:magic_angle}, at $\theta=3^{\circ}$ we are close to the magic angle for acoustic phonon absorption, leading to a particularly large Markov limit decay rate. While the decay rates (bottom) again show pronounced non-Markov\-ian dynamics via oscillations with a period $T\approx 1.25$~ps, the Markov limit is reached on a much faster time scale of $\approx 10$~ps, compared to $\approx 50$~ps in the case of $\theta=1^{\circ}$ (a, bottom). The oscillation period corresponds to the van Hove peak in the gPSD at $\hbar\Omega=\hbar 2\pi/T\approx 3.3$~meV [see Fig.~\ref{fig:gpsddisp_ac} (b)], which shifts to higher energy mismatches when increasing the twist angle, leading here to faster oscillations of the time dependent decay rate.

At sufficiently large twist angles of $\theta=5^{\circ}$ in Fig.~\ref{fig:dynamics_ac_4K} (c) the polarization decay is well described by the Markov limit (top) and the time dependent decay rate reaches the Markov limit on a much faster time scale of $\approx 3$~ps (bottom). It again shows non-Markov\-ian oscillations, which are significantly faster, since the corresponding van Hove peak in the gPSD in Fig.~\ref{fig:gpsddisp_ac} (c) lies at $\hbar\Omega\approx 20$~meV, corresponding to an oscillation period of $T\approx 200$~fs.

Overall, we see that the pho\-non-assisted dynamics of the absolute value of the moiré exciton polarization $|p_1(t)|$ are dominated by non-Markov\-ian effects at small twist angles, i.e., flat bands, and around the magic angle for acoustic phonon scattering. The former case corresponds closely to the independent boson model that is known to accurately describe excitons in quantum dots or at localized atomic defects~\cite{mahan2000many,krummheuer2002theory,nazir2016modelling,wigger2019phonon,preuss2022resonant}. At sufficiently large twist angles, i.e., curved bands corresponding to delocalized excitons~\cite{christiansen2017phonon,niehues2018strain,lengers2020theory,brem2020tunable,knorr2022exciton}, the absolute value of the polarization dynamics can be approximated as Markov\-ian without making a large error. However, as can be seen in Eqs.~\eqref{eq:pol_1}, the absolute value of the polarization dynamics does not contain all information on non-Markov\-ian effects, since its phase dynamics, i.e., the frequency components, are impacted by the imaginary part of the non-Markov\-ian dephasing function $\phi_1(t)$. This impact is best discussed in terms of PSBs of absorption spectra.

\subsubsection{Absorption spectra}

\begin{figure}
	\centering
	\includegraphics[width=\linewidth]{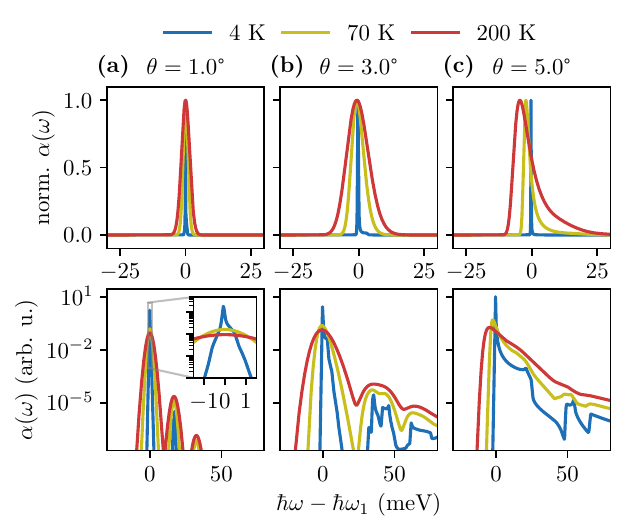}
	\caption{Absorption spectra of the lowest lying moiré exciton band including only acoustic phonon scattering for the twist angles $\theta=1^{\circ}$ (a), $\theta=3^{\circ}$ (b), and $\theta=5^{\circ}$ (c) and the temperatures $T=4$~K (blue), $T=70$~K (yellow), and $T=200$~K (red). Top: Normalized spectra on a linear scale. Bottom: Non-normalized spectra on a logarithmic scale.}
	\label{fig:absSpec_ac}
\end{figure}

Finally, we show the impact of acoustic pho\-nons on absorption spectra of the lowest lying moiré exciton band in Fig.~\ref{fig:absSpec_ac}. To this aim, we plot the spectra on a linear scale, normalized to their respective maximum in the top row, as well as on a logarithmic scale without normalization in the bottom row. We again consider the three twist angles $\theta=1^{\circ}$ (a), $\theta=3^{\circ}$ (b), and $\theta=5^{\circ}$ (c) and three different temperatures $T=4$~K (blue), $T=70$~K (yellow), and $T=200$~K (red). 

Starting the discussion with the normalized spectra on a linear scale [Fig.~\ref{fig:absSpec_ac} (top row)], we see that the spectra contain a single dominant peak, whose shape and position depend on temperature and twist angle. The position of the absorption peak generally shifts towards smaller energies with increasing temperature. This trend is well visible for $\theta=5^{\circ}$ (c) and becomes less visible for smaller twist angles. At $\theta=1^{\circ}$ (a) the peak virtually remains at the same position. The shift stems from the renormalization of the moiré exciton frequency $\omega_1\rightarrow\tilde{\omega}_1$ in Eqs.~\eqref{eq:pol_1} and corresponds to the polaron shift due to the dressing of the moiré excitons with pho\-nons. This polaron shift is temperature independent in the independent boson model, i.e., for localized excitons~\cite{mahan2000many,krummheuer2002theory,nazir2016modelling,wigger2019phonon,preuss2022resonant}. This explains why it is virtually temperature independent here for flat bands at $\theta=1^{\circ}$, due to the correspondence between flat bands and localized exciton systems discussed earlier. For larger twist angles the polaron shift obtains a pronounced temperature dependence which can also be observed for delocalized excitons, e.g., in TMDC monolayers~\cite{christiansen2017phonon,lengers2020theory}.

As discussed in the context of Eqs.~\eqref{eq:pol_1}, without any non-Markov\-ian pho\-non effects, the absorption spectrum would contain only a single Lorentzian, i.e., the ZPL, at $\omega\approx\omega_1$. Any deviations from such a single Lorentzian, especially asymmetries in the form of PSBs, are then an indication of non-Markov\-ian dynamics. In the case of the small twist angle $\theta=1^{\circ}$ (a, top), the absorption spectrum is essentially symmetric for all temperatures. While one might therefore interpret the corresponding spectrum as containing a single ZPL at elevated temperatures, in fact the symmetric peak stems from a broad PSB due to acoustic pho\-nons and the ZPL is completely suppressed. This can be seen especially well in the inset of Fig.~\ref{fig:absSpec_ac} (a, bottom). There we can see that for low temperatures (blue) we have a sharp ZPL on top of an asymmetric PSB background. This background becomes symmetric at elevated temperatures (yellow, red) since pho\-non emission and absorption processes involving acoustic pho\-nons with sufficiently small frequencies then become equally likely and the sharp ZPL is completely suppressed due to the increased number of pho\-nons interacting efficiently with the moiré exciton. This effect and the presence of symmetric PSBs at elevated temperatures are well established in the context of localized excitons coupling to acoustic pho\-nons, e.g., in quantum dots or color centers~\cite{krummheuer2002theory, wigger2019phonon}. Note that while the absorption spectrum becomes essentially symmetric with temperature on a linear scale (a, top), on the logarithmic scale (a, bottom), we find additional peaks at $\hbar(\omega-\omega_{1})\approx 17$~meV and $\approx 35$~meV. This stems from pho\-non-assisted transitions to higher lying bands, leading to additional peaks in the corresponding gPSD in Fig.~\ref{fig:gpsddisp_ac} (a). Here, we can therefore see that the shape of the gPSD determines the shape of the PSB, as derived in Eq.~\eqref{eq:PSB_1}.

If we now increase the twist angle to $\theta=3^{\circ}$ in Fig.~\ref{fig:absSpec_ac} (b, top), the absorption spectrum becomes slightly more asymmetric. At $T=4$~K (blue) we find an asymmetric PSB even on the linear scale (top) due to the non-Markovian polarization dynamics close to the magic angle [see Fig.~\ref{fig:dynamics_ac_4K} (b)], which however lead to less damping and therefore less broadening than the corresponding Markov limit, such that the asymmetric PSB is not masked by a broad ZPL.

The slight asymmetry, which is also present at elevated temperatures (yellow, red) can be attributed to the fact that the corresponding moiré exciton band is now slightly curved and pho\-non-assisted transitions are mainly possible for positive energy mismatches [Fig.~\ref{fig:gpsddisp_ac} (b)]. Due to the broader moiré exciton bandwidth at larger twist angles, the corresponding gPSD is broader, leading to a broader PSB attached to the ZPL, as can be seen on the logarithmic scale Fig.~\ref{fig:absSpec_ac} (b, bottom) at low temperatures (blue). We again find additional peaks in the absorption spectrum on a logarithmic scale above $\hbar(\omega-\omega_1)\approx 30$~meV stemming from transitions to higher lying bands.

Increasing the twist angle even further to $\theta=5^{\circ}$ in Fig.~\ref{fig:absSpec_ac} (c) increases the asymmetry of the spectra especially at elevated temperatures (yellow, red), stemming from the even broader gPSD [Fig.~\ref{fig:gpsddisp_ac} (c)] due to the increased bandwidth of the lowest-lying moiré exciton band. At $T=4$~K (blue) we can identify a small peak in the PSB at $\hbar(\omega-\omega_{1})\approx 20$~meV on the logarithmic scale stemming from transitions to the van Hove singularity at the $m$-point. Increasing the temperature (yellow, red) leads to a single dominant asymmetric peak, washing out the details of the PSB. Such asymmetric absorption spectra for large twist angles have been reported for moiré excitons in Ref.~\cite{jin2019observation}, however they have not been attributed to acoustic phonon scattering therein. This asymmetry is also well known in the context of delocalized excitons, e.g., in TMDCs~\cite{christiansen2017phonon,lengers2020theory}, showing again that the case of sufficiently large twist angles can be understood in terms of delocalized excitons.


We can summarize the findings in this section on the impact of acoustic pho\-nons as follows: 

(i) For sufficiently small twist angles in the flat-band regime, the moiré exciton-pho\-non interaction is dominated by non-Markov\-ian effects. In this regime, the system exhibits features well-known from the independent boson model that is used to describe pho\-nons coupling to localized excitons in quantum dots or color centers~\cite{mahan2000many,krummheuer2002theory,nazir2016modelling,wigger2019phonon,preuss2022resonant}. 

(ii) For sufficiently large twist angles the moiré exciton bands are curved. The essential features, e.g., asymmetric absorption peaks at elevated temperatures, are similar to the case of delocalized excitons, e.g., in TMDC monolayers~\cite{christiansen2017phonon,lengers2020theory}.

\subsection{Influence of optical pho\-nons}
\subsubsection{gPSD for optical pho\-non scattering}
\begin{figure}
	\centering
	\includegraphics[width=\linewidth]{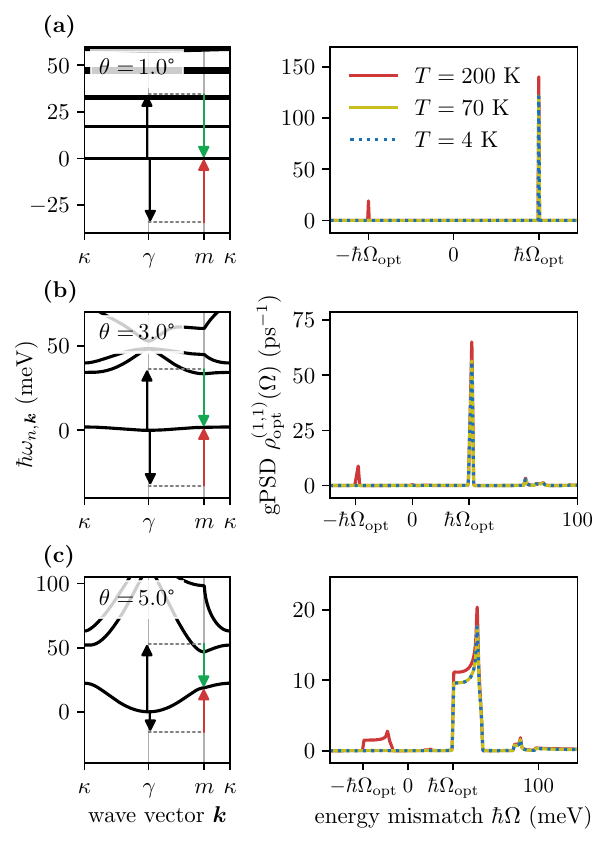}
	\caption{Moiré exciton band structure $\hbar\omega_{n,\bm{k}}$ (left) and corresponding gPSD $\rho_{j=\rm{opt}}^{(1,1)}(\Omega)$ of the lowest lying moiré exciton band due to optical pho\-non scattering (right) for the three twist angles $\theta=1^{\circ}$ (a), $3^{\circ}$ (b), $5^{\circ}$ (c) and three different temperatures $T=4$~K (blue), $T=70$~K (yellow), and $T=200$~K (red). Optical pho\-non-assisted intra\-band transitions from the $\gamma$- to the $m$-point of the lowest lying moiré exciton band (left) are sketched for phonon absorption (red arrows) and emission (green arrows) with black vertical arrows corresponding to the energy mismatch $\hbar\Omega$ that is needed for that transition to occur.}
	\label{fig:gpsddisp_opt}
\end{figure}
We now consider the impact of optical pho\-nons on the moiré excitons. To this aim, in Fig.~\ref{fig:gpsddisp_opt} we show the gPSD $\rho_{j=\rm{opt}}^{(1,1)}(\Omega)$ of the lowest lying moiré exciton band due to optical pho\-non scattering as a function of the energy mismatch $\hbar\Omega$. From the top to bottom row, we consider the three twist angles $\theta=1^{\circ}$ (a), $\theta=3^{\circ}$ (b), and $\theta=5^{\circ}$ (c). The left column shows the respective moiré exciton band structure, which was already presented in Fig.~\ref{fig:band_structure}. In addition, optical pho\-non-assisted intra\-band transitions of the lowest lying moiré exciton band are sketched in the following way: The absorption (emission) of an optical pho\-non is drawn as a red upward (green downward) arrow with the length corresponding to the energy $\hbar\Omega_{j=\rm{opt}}$ of the optical pho\-non mode. The black vertical arrows represent the energy mismatch $\hbar\Omega$ necessary for that transition to occur and the dashed horizontal line represents the wave vector of the optical pho\-non contributing to the transition. We specifically show the pho\-non-in\-duced transitions from the $\gamma$-point (the optically active moiré exciton) to the $m$-point, i.e., the van Hove singularity of the lowest lying moiré exciton band. These transitions dominate the gPSDs shown in the right column of Fig.~\ref{fig:gpsddisp_opt} analogous to the case of acoustic pho\-nons in Fig.~\ref{fig:gpsddisp_ac}. For the presented gPSDs we consider three different temperatures $T=4$~K (blue), $T=70$~K (yellow), and $T=200$~K (red).

Starting with a small twist angle of $\theta=1^{\circ}$ in Fig.~\ref{fig:gpsddisp_opt} (a), we see a dominant peak in the gPSD hat $\Omega=\Omega_{\rm{opt}}$. This peak stems from optical pho\-non emission processes from the $\gamma$-point of the lowest-lying band into the same band. Since the bands are flat at such small twist angles, we get a very narrow peak. When increasing the temperature sufficiently (red, 200K), an additional pho\-non absorption peak at $\Omega=-\Omega_{\rm{opt}}$ appears. 


\begin{figure}
	\centering
	\includegraphics[width=\linewidth]{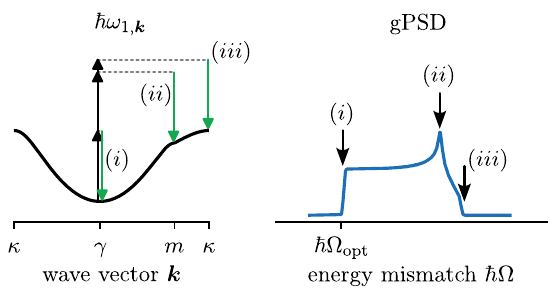}
	\caption{Schematic of the lowest lying moiré exciton band with intra\-band transitions from the $\gamma$-point to high symmetry points induced by optical phonon emission (left). Schematic of the corresponding gPSD (right).}
	\label{fig:gpsd_characteristicShape}
\end{figure}

For larger twist angles $\theta=3^{\circ}$ (b) and $\theta=5^{\circ}$ (c) the narrow peak is broadened towards a characteristic shape lying above $\Omega\geq \Omega_{\rm{opt}}$ for pho\-non emission processes and $\Omega\geq -\Omega_{\rm{opt}}$ for pho\-non absorption processes. In addition we find small contributions to the gPSD due to inter\-band scattering, not discussed in the following.

We can understand the characteristic shape in the gPSD with the help of the schematic presented in Fig.~\ref{fig:gpsd_characteristicShape}. It shows the typical optical pho\-non emission processes in the lowest lying moiré band starting from the $\gamma$-point (left) and the corresponding schematic gPSD (right) for the region $\Omega\geq \Omega_{\rm{opt}}$. The smallest possible energy mismatch due to pho\-non emission processes stems from a transition from the $\gamma$-point to the $\gamma$-point (i), leading to the steep low energy flank of the gPSD at $\Omega=\Omega_{\rm{opt}}$. The largest possible energy mismatch stems from a transition to the $\kappa$-point (iii), leading to a cutoff of the gPSD at $\Omega=\omega_{1,\bm{k}=\kappa}-\omega_{1,\bm{k}=\gamma}+\Omega_{\rm{opt}}$. In between these two situations we find the transition to the $m$-point (ii), i.e., the van Hove singularity of the lowest-lying moiré exciton band, leading to a  peak in the gPSD due to the high density of states. 

The gPSD for optical pho\-non scattering is effectively given by the density of states of the moiré exciton band, which becomes clear when considering that Eq.~\eqref{eq:g_PSD_1} for $j=\mathrm{opt}$ and $n'=1$ can be written as
\begin{equation}\label{eq:gPSD_density_of_states}
	\rho_{j=\rm{opt}}^{(1,1)}(\Omega)\sim \sum_{\bm{Q},\sigma=\pm}\delta(\Omega+\sigma\Omega_{\rm{opt}}-\omega_{1,-\bm{Q}})
\end{equation}
since the optical pho\-non mode is dispersion-less and we set $\omega_{1,\bm{0}}=0$. This is simply the density of states of the lowest-lying moiré exciton band, shifted by $\mp\Omega_{\rm{opt}}$ for pho\-non absorption (+) and emission ($-$) processes. For flat bands at small twist angles in Fig.~\ref{fig:gpsddisp_opt} (a) this leads to sharp peaks in the gPSD. An increase of the twist angle leads to a larger bandwidth of the lowest-lying moiré exciton band in (b) and (c). Therefore the density of states of the band, as well as the characteristic shape in the gPSD, become broadened accordingly.

As discussed in the following, this relation between the moiré exciton density of states and the gPSD explains the steep low energy flank at $\Omega=\pm\Omega_{\mathrm{opt}}$, since in this region only moiré excitons with $\bm{Q}\rightarrow\bm{0}$ contribute in Eq.~\eqref{eq:gPSD_density_of_states}. This means that we can approximate the relevant part of the moiré exciton dispersion relation harmonically with an effective mass $\overline{M}$, leading to the gPSD
\begin{equation}
	\rho_{j=\rm{opt}}^{(1,1)}(\Omega\mp\Omega_{\mathrm{opt}})\sim \sum_{\bm{Q}}\delta\qty(\Omega-\frac{\hbar\bm{Q}^2}{2\overline{M}})\sim \Theta(\Omega)\,.
\end{equation}
We get the steep low energy flanks in Figs.~\ref{fig:gpsddisp_opt} and \ref{fig:gpsd_characteristicShape}, since the density of states for a quadratic dispersion relation in two dimensions has the shape of a Heaviside function $\Theta(\Omega)$. 

Note that we already found a similar behavior for acoustic pho\-nons in Fig.~\ref{fig:gpsddisp_ac}. Especially at large twist angles, the gPSD for acoustic pho\-non scattering in Fig.~\ref{fig:gpsddisp_ac} has a characteristic shape similar to the one presented in Fig.~\ref{fig:gpsd_characteristicShape}. We can understand this now analogously to Eq.~\eqref{eq:gPSD_density_of_states}. If we approximate the acoustic pho\-non energy by $\Omega_{\mathrm{ac},\bm{Q}}\approx 0$ in Eq.~\eqref{eq:g_PSD_1} [the arrows in Fig.~\ref{fig:gpsddisp_ac} (c) are quasi horizontal compared to the moiré exciton dispersion for large twist angles], we obtain Eq.~\eqref{eq:gPSD_density_of_states} but with $\Omega_{\rm{opt}}\rightarrow 0$, i.e., simply the density of states of the moiré excitons. This approximation does not account for the two van Hove peaks visible in Fig.~\ref{fig:gpsddisp_ac}, which are due to the small energy difference between pho\-non absorption and emission processes. This splits the single peak in the characteristic shape of Fig.~\ref{fig:gpsd_characteristicShape} in two in the gPSD from Fig.~\ref{fig:gpsddisp_ac} (c).

Note that for all considered cases in Fig.~\ref{fig:gpsddisp_opt} the optical pho\-non gPSD vanishes at $\Omega=0$, such that the coupling with optical pho\-nons typically does not lead to any relevant Markov limit decay rate in the single-band system. Analogous to Eq.~\eqref{eq:magic_angle}, we find a magic angle for optical pho\-non scattering, where optical pho\-non absorption induces energy conserving transitions from the $\gamma$- to the $m$-point, at $\theta_c=6.1^{\circ}$. At this magic angle for optical pho\-non absorption we can only have a significant impact on the Markov limit decay rate at elevated temperatures $k_{\mathrm{B}}T\sim \hbar\Omega_{\rm{opt}}$ (here around 200~K, see Fig.~\ref{fig:gpsddisp_opt}).

\subsubsection{Influence of both pho\-non branches on absorption spectra of the lowest-lying band}

\begin{figure}
	\centering
	\includegraphics[width=\linewidth]{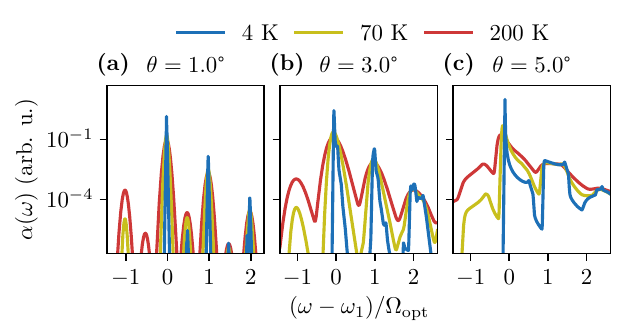}
	\caption{Absorption spectra on a logarithmic scale of the lowest lying moiré exciton band including the impact of both acoustic and optical phonon scattering for the twist angles $\theta=1^{\circ}$ (a), $\theta=3^{\circ}$ (b), and $\theta=5^{\circ}$ (c) and the temperatures $T=4$~K (blue), $T=70$~K (yellow), and $T=200$~K (red).}
	\label{fig:absSpec_both}
\end{figure}

After investigating the influence of optical pho\-nons on the gPSD in the previous section, we conclude the discussion on the optical properties of the lowest-lying moiré exciton band by presenting absorption spectra, including the impact of both pho\-non bran\-ches, in Fig.~\ref{fig:absSpec_both}. We consider the three twist angles $\theta=1^{\circ}$ (a), $\theta=3^{\circ}$ (b), and $\theta=5^{\circ}$ (c) and three different temperatures $T=4$~K (blue), $T=70$~K (yellow), and $T=200$~K (red).

If we compare these spectra directly with the ones shown in Fig.~\ref{fig:absSpec_ac}, we can identify the additional influence of the optical pho\-nons. Since optical pho\-nons do not impact the Markov\-ian dynamics here (gPSDs in Fig.~\ref{fig:gpsddisp_opt} vanish at $\Omega=0$), their main influence lies in generating additional PSBs at $\omega-\omega_1\approx n\Omega_{\rm{opt}}$. These are visible as narrow peaks in the case of flat bands (a) and the width of these peaks gets larger when increasing the twist angle (b, c), in accordance with the discussion on the gPSD due to optical pho\-non scattering in  the context of Figs.~\ref{fig:gpsddisp_opt} and \ref{fig:gpsd_characteristicShape}. Note that the spectra in Fig.~\ref{fig:absSpec_both} (a) contain multiple additional peaks stemming from pho\-non-induced transitions to higher lying bands at $\omega-\omega_1\neq n\Omega_{\rm{opt}}$.

The presence of narrow PSBs due to optical pho\-nons in Fig.~\ref{fig:absSpec_both} (a) is well known from the study of localized excitons in quantum dots and color centers~\cite{mahan2000many,krummheuer2002theory,wigger2019phonon,preuss2022resonant}, again emphasizing the correspondence between small twist angles, i.e., flat bands, and localized excitons. The PSBs due to optical phonon scattering lie at $\omega-\omega_1\approx n\Omega_{\mathrm{opt}}$ which shows that the TCL approach includes multi-pho\-non processes, e.g., the emission of two optical pho\-nons during the light absorption process leading to a PSB at $\omega-\omega_1\approx 2\Omega_{\mathrm{opt}}$. At elevated temperatures (yellow, red), we furthermore observe an optical pho\-non absorption peak at $\omega-\omega_1\approx -\Omega_{\rm{opt}}$. 

Increasing the twist angle [Fig.~\ref{fig:absSpec_both} (b, c)] leads to curved bands, a larger bandwidth inducing a broader gPSD (see Fig.~\ref{fig:gpsddisp_opt}), and thus eventually to partially overlapping optical PSB peaks, which are however clearly separated from the ZPL for low temperatures (blue) by the optical pho\-non frequency $\Omega_{\rm{opt}}$. At elevated temperatures (yellow, red) the ZPL merges with the optical PSBs. This closely resembles the influence of optical pho\-nons on the absorption spectra of delocalized excitons, e.g., in TMDCs~\cite{christiansen2017phonon,lengers2020theory}.

\section{Twist angle dependence of the multi-band absorption spectrum}
\subsection{Twist angle dependence of the dipole moments}
\begin{figure}
	\centering
	\includegraphics[width=0.9\linewidth]{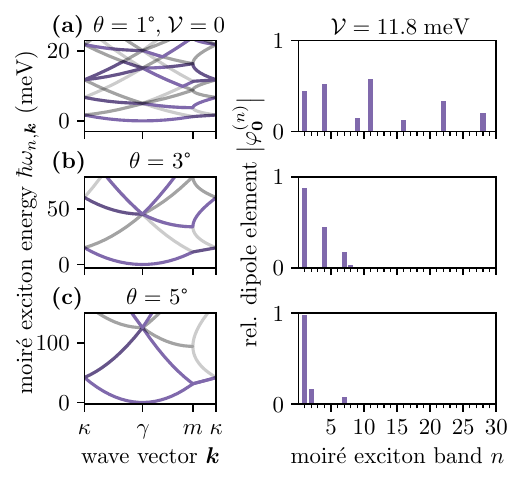}
	\caption{Twist angle dependence of the dipole matrix elements. Left: Moiré exciton band structure for a vanishing moiré potential $\mathcal{V}=0$. The colored bands are those which obtain a non-vanishing dipole matrix element at the $\gamma$-point when switching on the moiré potential to $\mathcal{V}=11.8$~meV, as can be seen on the right.}
	\label{fig:dipole_MBZtutorial}
\end{figure}
In the following we will discuss the absorption spectrum of the moiré exciton system including multiple bands, instead of just the lowest lying one as in the previous section. In absence of exciton-pho\-non scattering we can solve Eq.~\eqref{eq:TCL} to obtain the absorption spectrum in Eq.~\eqref{eq:alpha_omega} as
\begin{equation}\label{eq:alpha_no_phon}
	\alpha(\omega)\sim \sum_n |M_n|^2\frac{\gamma_n/2}{(\gamma_n/2)^2+(\omega-\omega_n)^2}\,.
\end{equation}
This is just a sum of Lorentz\-ians for each moiré ex\-citon at the $\gamma$-point, weighted by the absolute square of their respective dipole moment $M_n$. The maxima of these peaks are given by $\sim |M_n|^2/\gamma_n$ and thus twist angle independent according to Eqs. \eqref{eq:def_M_n} and \eqref{eq:gamma_n}, since the factor $|\varphi_{\bm{0}}^{(n)}|$, i.e., the relative dipole moment of the moiré excitons at the $\gamma$-point, cancels. The absorption spectrum in absence of any pho\-non influence would therefore contain peaks with identical height for each moiré band. This is clearly in contrast to physical intuition which tells us that brighter moiré excitons, i.e., with larger dipole moment $|M_n|$ should lead to more dominant absorption peaks. Note however that the area under each peak is given by $\sim |M_n|^2$ and therefore larger for brighter moiré excitons and vanishing for dark ones. In any actual measurement of the absorption spectrum a spectrometer with finite resolution would be involved, leading to a convolution of Eq.~\eqref{eq:alpha_no_phon} with the spectrometer response. Assuming a Lorentzian spectrometer broadening $\Gamma$ modeled via~\cite{eberly1977time, preuss2022resonant,groll2025fundamentals}
\begin{equation}\label{eq:spectrometer}
	p_n(t)\rightarrow p_n(t) e^{-\Gamma t}
\end{equation}
changes Eq.~\eqref{eq:alpha_no_phon} into
\begin{equation}\label{eq:alpha_no_phon_Gamma}
	\alpha(\omega)\sim \sum_n |M_n|^2\frac{\gamma_n/2+\Gamma}{(\gamma_n/2+\Gamma)^2+(\omega-\omega_n)^2}\,.
\end{equation}
Especially for $\gamma_n\ll \Gamma$, the maximum of the absorption peaks is now determined by $|M_n|^2/\Gamma$ and not $|M_n|^2/\gamma_n$ and therefore scales directly with the absolute square of the relative dipole moments $|\varphi_{\bm{0}}^{(n)}|^2$ of the moiré excitons. Brighter moiré excitons then lead to larger peaks and darker ones to smaller peaks. This discussion demonstrates that it is sometimes crucial to consider the impact of an actual detection process when interpreting simulated spectroscopy signals.

Since the dipole moment $M_n$ is determined by the moiré exciton wavefunction at the $\gamma$-point $\varphi_{\bm{0}}^{(n)}$ [see Eq.~\eqref{eq:def_M_n}], it is twist angle dependent. To understand this dependence, Fig.~\ref{fig:dipole_MBZtutorial} shows the absolute value of the relative dipole moments $|\varphi_{\bm{0}}^{(n)}|$ for several moiré exciton bands in the right column for the twist angles $\theta=1^{\circ}$ (a), $\theta=3^{\circ}$ (b), and $\theta=5^{\circ}$ (c). We can see that the number of bands which carry a non-vanishing dipole moment decreases with increasing twist angle. To better understand this, in the left column we consider the moiré exciton band structures for the respective twist angles for a vanishing moiré potential $\mathcal{V}=0$. The colored bands are those which obtain a non-vanishing dipole moment at the $\gamma$-point when switching on the moiré potential, using our standard parameter $\mathcal{V}=11.8$~meV, i.e., correspond to peaks in the respective figures in the right column.

According to Eq.~\eqref{eq:eigenvalue_moire} in absence of the moiré potential the moiré exciton dispersion relation displayed in Fig.~\ref{fig:dipole_MBZtutorial} is simply the quadratic dispersion relation of the homogeneous exciton, folded back into the first MBZ. A non-vanishing moiré potential in Eq.~\eqref{eq:eigenvalue_moire} would couple homogeneous exciton states at the same position in the first MBZ. Since only the homogeneous exciton with $\bm{K}=\bm{0}$ is bright [see Eq.~\eqref{eq:V_ex_laser_original}], only moiré excitons at the $\gamma$-point can gain a non-vanishing dipole moment via this moiré potential coupling [see Eq.~\eqref{eq:V_ex_laser}]. As can be seen in Fig.~\ref{fig:dipole_MBZtutorial} (left), the homogeneous exciton states at the $\gamma$-point lie closer to the lowest-lying bright one for smaller twist angles (a) compared to larger twist angles (b, c). This is due to the fact that the MBZ is smaller for smaller twist angles. If we now couple these unperturbed homogeneous exciton states which are folded back into the first MBZ with a non-vanishing moiré potential $\mathcal{V}\neq 0$, the coupling is stronger, the closer they are in energy. For this reason, more moiré exciton states become bright for smaller twist angles in Fig.~\ref{fig:dipole_MBZtutorial} (right), inheriting their non-vanishing dipole moment from the lowest lying unperturbed homogeneous exciton at the $\gamma$-point~\cite{jin2019observation, brem2020tunable,huang2022excitons}.

While this discussion explains the overall 'envelope' of the dipole elements observed in Fig.~\ref{fig:dipole_MBZtutorial} (a), it does not explain why the majority of bands is dark for all of the displayed twist angles. This can be understood from the symmetry of the system. As discussed in detail in App.~\ref{app:symmetries}, we can classify the solutions to Eq.~\eqref{eq:eigenvalue_moire} at the $\gamma$-point of the MBZ via the $C_{3v}$ point group symmetry of the moiré potential, which originates from the $C_{3v}$ point group symmetry of the TMDC monolayers. We find that the bright moiré exciton states belong to the one-dimensional irreducible representation (irrep) $A_1$, while dark states belong either to the two-dimensional irrep $E$ or the one-dimensional irrep $A_2$~\cite{li1985solution,ruiz2020theory}. This explains why the majority of moiré exciton bands is dark for all displayed twist angles in Fig.~\ref{fig:dipole_MBZtutorial}.

\subsection{Suppression of moiré exciton peaks via efficient optical pho\-non emission}

\begin{figure}
	\centering
	\includegraphics[width=0.9\linewidth]{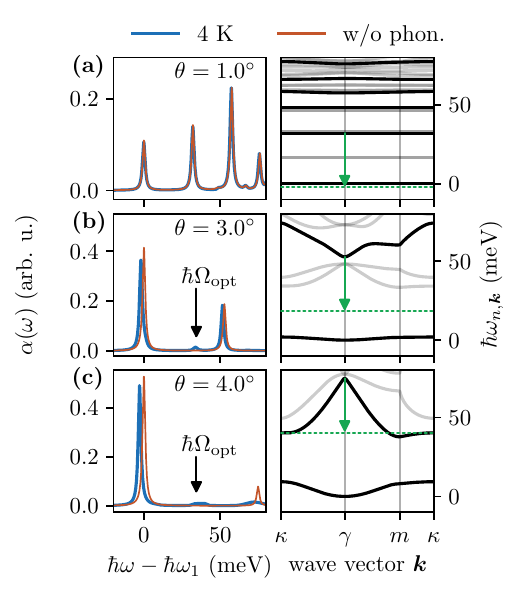}
	\caption{Absorption spectra (left) and dispersion relations (right) for the twist angles $\theta=1^{\circ}$ (a), $\theta=3^{\circ}$ (b), and $\theta=4^{\circ}$ (c). Absorption spectra are calculated for the full multi-band system, solving Eq.~\eqref{eq:TCL} without inter-polarization coupling, at $T=4$~K (blue) and without coupling to phonons (orange). We include a spectrometer resolution of $\hbar\Gamma=1$~meV via Eq.~\eqref{eq:spectrometer}. The dispersion relations show the bright moiré bands, i.e., those containing bright excitons at the $\gamma$-point (black), as well as dark bands (grey). Optical phonon emission starting in the second bright moiré band (green arrow) leads to energy conserving intra\-band transitions if the horizontal green line intersects with the second bright band.}
	\label{fig:absSpec_multipleBands}
\end{figure}
After this discussion on the twist angle dependence of moiré exciton dipole moments, we will focus on the absorption spectrum of the moiré exciton system with multiple bands, including the influence of both, acoustic and optical, pho\-non branches. This implies that we have to solve the full TCL master equation~\eqref{eq:TCL}. In the following we will however neglect any inter-polarization coupling with $n''\neq n$ in Eq.~\eqref{eq:TCL}. While we found numerically that this coupling typically does not play a significant role, in App.~\ref{app:inter_pol} we support this finding by a discussion on analytical properties of the theory. Note however that neglecting the inter-polarization coupling does not mean that we neglect pho\-non-in\-duced scattering between moiré bands, since these are contained in the gPSD in Eq.~\eqref{eq:g_PSD} also for the case $n''=n$ via the sum over $n'$. Without inter-polarization coupling the coupled TCL master equation~\eqref{eq:TCL} simplifies to a set of decoupled equations for each moiré band. Since only bright bands with $M_n\neq 0$ are excited optically and contribute to the absorption spectrum [see Eqs.~\eqref{eq:V_ex_laser} and \eqref{eq:alpha_omega}], this simplifies the numerical simulation significantly and we can focus on these bright bands, calculating only the corresponding gPSDs in Eq.~\eqref{eq:g_PSD} for $n''=n$.

Figure~\ref{fig:absSpec_multipleBands} shows the absorption spectrum of the multi-band system (left column) without inter-polari\-zation coupling for $T=4$~K and the twist angles $\theta=1^{\circ}$ (a), $\theta=3^{\circ}$ (b), and $\theta=4^{\circ}$ (c) from top to bottom. We show the full simulation including the impact of pho\-nons (blue), as well as the spectra when neglecting exciton-pho\-non coupling (orange). In this fashion we can identify which part of the twist angle dependence is due to pho\-non scattering and which part is due to the behavior of the dipole moments from Fig.~\ref{fig:dipole_MBZtutorial}. In all spectra we include a spectrometer broadening of $\hbar\Gamma=1$~meV [see Eq.~\eqref{eq:spectrometer}]. In addition to the absorption spectra, for each twist angle we show the moiré exciton band structure (right column) with bright bands, i.e., those containing bright moiré excitons at the $\gamma$-point, in black, and all other bands in grey.

Focusing on the case without exciton-pho\-non coupling first (orange), we see that the absorption spectra in Fig.~\ref{fig:absSpec_multipleBands} consist of peaks at the bright moiré exciton energies $\hbar\omega_n=\hbar\omega_{n,\bm{k}=\gamma}$ (black, right) whose height is approximately given by $|M_n|^2$ according to Eq.~\eqref{eq:alpha_no_phon_Gamma}. The peak height thus depends on the twist angle via the dipole moments (see Fig.~\ref{fig:dipole_MBZtutorial}). 

If we include the coupling to pho\-nons (blue), we observe three effects: (i) The position of the peaks shifts towards smaller energies due to the polaron shift, (ii) additional peaks at $\hbar\omega-\hbar\omega_1\approx\hbar\Omega_{\mathrm{opt}}$, i.e., optical PSBs, appear, and (iii) the intensities of the original peaks change. All of these features are twist-angle dependent. For the smallest twist angle $\theta=1^{\circ}$ (a) the inclusion of phonons has virtually no impact due to the comparatively large spectrometer broadening $\hbar\Gamma=1$~meV included in the simulations. Increasing the twist angle (b, c) leads to a visible polaron shift [effect (i)], whose strength grows with the twist angle. Also effect (ii), i.e., the presence of the optical PSB at $\hbar\omega-\hbar\omega_1\approx \hbar\Omega_{\rm{opt}}$ is only visible for increased twist angles (b, c) and the PSB gets broader when increasing the twist angle, as discussed in the context of Figs.~\ref{fig:gpsd_characteristicShape} and \ref{fig:absSpec_both}. At  $\theta=1^{\circ}$ (a) there is a large bright moiré exciton peak at the energy where we would expect the optical PSB, such that it is not explicitly visible there.

Effect (iii), i.e., the modification of the intensity of the peaks, depends strongly on the twist angle. On the one hand, at the smaller twist angles of $\theta=1^{\circ}$ and $\theta=3^{\circ}$ in Fig.~\ref{fig:absSpec_multipleBands} (a, b) all peaks remain visible when including the coupling to phonons. However they are potentially broadened leading to a decrease in peak height, especially for the lowest lying bright moiré exciton at $\theta=3^{\circ}$ (b). For the largest twist angle of $\theta=4^{\circ}$ (c) on the other hand the coupling to phonons leads to a complete suppression of the absorption peak belonging to the second bright moiré exciton band.

To understand this abrupt transition from an absorption spectrum containing multiple ZPLs for sufficiently small twist angles to just one ZPL for larger ones, the right column of Fig.~\ref{fig:absSpec_multipleBands} shows energy conserving intra\-band scattering processes in the second bright band due to optical phonon emission as green downward arrows starting at the respective $\gamma$-point. These are possible when the dashed horizontal line at the tip of the arrow intersects the second bright band. We see that for the twist angles $\theta=1^{\circ}$ (a) and $\theta=3^{\circ}$ (b) there are no such energy conserving intra\-band transitions via optical pho\-nons due to the insufficient bandwidth and unsuitable orientation ($\gamma$-point being a minimum) of the second bright band. For larger twist angles, here $\theta=4^{\circ}$ (c), the bandwidth surpasses the optical pho\-non energy $\hbar\Omega_{\rm{opt}}$ and the $\gamma$-point becomes a maximum such that efficient energy conserving transitions via emission of optical pho\-nons become possible, leading to a large Markov decay rate of the second bright moiré exciton band and a complete suppression of the corresponding absorption peak (left, blue vs. orange). This discussion demonstrates that pho\-non scattering can have a crucial impact when studying the optical properties of moiré excitons.

\section{Conclusion}
We have investigated the twist-angle dependent influence of exciton-phonon coupling on the optical properties of intralayer excitons in a twisted TMDC bilayer. Starting from a two-band model for electrons and holes we derived a Hamiltonian describing the interaction of moiré excitons with pho\-nons and an external light field. We then derived a time-convolution\-less (TCL) master equation for the microscopic moiré exciton polarizations whose dynamics after an ultrashort light pulse can be used to calculate the linear absorption spectrum. We found that the generalized phonon spectral density (gPSD) is especially helpful for understanding the influence of pho\-nons on the optical properties of moiré excitons. We applied our theory to the specific example of MoSe$_2$ intralayer excitons in a twisted MoSe$_2$/ WSe$_2$ hetero-bilayer for twist angles in the range of $1^{\circ}$ to $5^{\circ}$. 

Focusing first on the optical properties of the lowest lying moiré exciton band we investigated separately the influence of acoustic and optical pho\-nons. In the case of acoustic pho\-nons we found two distinct regimes of exciton-phonon coupling: (i) At small twist angles where the moiré exciton bands are flat the decay of the microscopic polarization is dominated by non-Markovian dynamics and the absorption spectra resemble those found in localized exciton systems, e.g., quantum dots or color centers. (ii) At sufficiently large twist angles the polarization decay is approximately Markovian while the absorption spectra contain a strongly asymmetric peak, similar to experimental reports~\cite{jin2019observation}. This resembles the line shape found in monolayer TMDCs, i.e., for delocalized excitons, where it is also attributed to acoustic phonon scattering. In between these two limiting cases lies the magic angle for acoustic phonon scattering, where the coupling between moiré excitons in the lowest lying band and acoustic pho\-nons increases drastically.

Optical pho\-nons impact the absorption spectrum of the lowest lying moiré exciton band mainly by introducing optical phonon sidebands (PSBs). These are narrow for small twist angles, i.e., flat moiré exciton dispersions with small bandwidths, and become broader for larger twist angles, i.e., moiré exciton dispersions with larger bandwidths. Using the properties of the gPSD, we found a direct connection between the shape of these PSBs and the density of states of the moiré exciton band that they belong to, which explains this behavior.

In the full multi-band system we first investigated the twist angle dependence of the moiré exci\-ton di\-pole moments. Including then the interaction with pho\-nons we found that the absorption spectra are influenced critically by intra\-band scattering due to optical pho\-nons, suppressing absorption peaks of higher lying bands once the corresponding twist angle dependent bandwidth surpasses the optical phonon energy. This discovery implies that in addition to the twist angle dependence of the moiré exciton dipole moments~\cite{jin2019observation} one might need to consider moiré ex\-citon-pho\-non scattering in discussions of the optical properties of moiré intralayer excitons.

\appendix
\section{Derivation of the exciton picture Hamiltonian}\label{app:exciton_picture}
We consider a two-band model in effective mass approximation, describing electrons and holes in a direct semiconductor and we ignore electron-electron and hole-hole interaction consistent with the assumption of small charge carrier densities discussed in the following~\cite{axt1994dynamics,katsch2018theory, lengers2020theory}. We furthermore suppress the spin index, assuming that the system is excited with a suitable polarization and energy and only intra\-layer excitons with a specific combination of electron and hole spins are excited due to spin-valley locking in TMDCs~\cite{xiao2012coupled, wang2018colloquium}. The Hamiltonian is given by
\begin{align}
	H_{\rm eh}=&\sum_{\bm{k}}\qty(\epsilon^{(\rm e)}_{\bm{k}} c_{\bm{k}}^{\dagger}c_{\bm{k}}^{}+\epsilon_{\bm{k}}^{(\rm h)}d_{\bm{k}}^{\dagger}d_{\bm{k}}^{})\notag\\
	&-\sum_{\bm{k}\bm{k}'\bm{q}}v(\bm{q})c_{\bm{k}}^{\dagger}d_{\bm{k}'}^{\dagger}d_{\bm{k}'-\bm{q}}^{}c_{\bm{k}+\bm{q}}^{}\label{eq:H_eh}
\end{align}
with the respective electron and hole energies
\begin{equation}
	\epsilon^{(\rm e)}_{\bm{k}}=\frac{\hbar^2\bm{k}^2}{2m_{\rm e}}+E_{\rm G}\,,\qquad \epsilon^{(\rm h)}_{\bm{k}}=\frac{\hbar^2\bm{k}^2}{2m_{\rm h}}\,.
\end{equation}
Here, $m_{\mathrm{e/h}}$ are the effective masses of electron and hole and $E_{\rm G}$ is the bandgap energy. $c_{\bm{k}}^{(\dagger)}$ and $d_{\bm{k}}^{(\dagger)}$ are the respective electron and hole annihilation (creation) operators. The Coulomb matrix element $v(\bm{q})$ can be considered arbitrary at this point. For TMDC monolayers and their corresponding heterostructures it can be modeled via the Rytova-Keldysh or related potentials~\cite{chernikov2014exciton,ovesen2019interlayer, brem2020phonon, lengers2020theory}. We construct the operators
\begin{equation}\label{eq:def_Y_dag}
	Y_{l,\bm{Q}}^{\dagger}=\sum_{\bm{k}}\Phi_l(\bm{k}-\mu_{\rm e}\bm{Q})c_{\bm{k}}^{\dagger}d_{\bm{Q}-\bm{k}}^{\dagger}
\end{equation}
describing the creation of an electron-hole pair with total momentum $\hbar\bm{Q}$. Here, $\mu_{\rm e}=m_{\rm e}/M$ is the electron mass fraction with respect to the total mass $M=m_{\rm e}+m_{\rm h}$ of the pair. Requiring that these electron-hole pairs are eigenstates of the two-band Hamiltonian including electron-hole interaction
\begin{equation}
	H_{\rm eh}Y_{l,\bm{Q}}^{\dagger}\ket{0}=E_{l,\bm{Q}}Y_{l,\bm{Q}}^{\dagger}\ket{0}
\end{equation}
with the electron-hole vacuum state $\ket{0}$, leads to the Wannier equation
\begin{equation}\label{eq:wannier}
	\sum_{\bm{q}}\qty[\frac{\hbar^2\bm{k}^2}{2\mu}\delta_{\bm{q}\bm{0}}-v(\bm{q})]\Phi_l(\bm{k}+\bm{q})=E_l\Phi_l(\bm{k})
\end{equation}
with the reduced mass $\mu=m_{\rm e}m_{\rm h}/M$ of the electron-hole pair. This equation determines the exciton binding energies $E_l$, which are connected to the total energy of the electron-hole pair, i.e., the exciton, as
\begin{equation}
	E_{l,\bm{Q}}=E_l+E_{\rm G}+\frac{\hbar^2\bm{Q}^2}{2M}\,.
\end{equation}
The exciton wavefunctions form a complete
\begin{equation}
	\sum_l \Phi^*_l(\bm{k}')\Phi_l^{}(\bm{k})=\delta_{\bm{k}\bm{k}'}
\end{equation}
and orthonormal set
\begin{equation}
	\sum_{\bm{k}} \Phi^*_l(\bm{k})\Phi_{l'}^{}(\bm{k})=\delta_{ll'}\,,
\end{equation}
being eigenfunctions of the Wannier equation. For this reason we can also invert the definition of the exciton creation operators in Eq.~\eqref{eq:def_Y_dag} to obtain
\begin{equation}\label{eq:invert_def_Y_dag}
	\sum_l \Phi_l^*(\bm{k}-\mu_{\rm e}\bm{Q})Y^{\dagger}_{l,\bm{Q}}=c_{\bm{k}}^{\dagger}d_{\bm{Q}-\bm{k}}^{\dagger}\,.
\end{equation}
Since here we are interested in the linear absorption spectrum, we can restrict ourselves to the subspace of zero and one exciton, in which case for a neutral semiconductor we can write
\begin{equation}\label{eq:low_density_e}
	c_{\bm{k}'}^{\dagger}c_{\bm{k}}^{}\approx N_{\rm h}c_{\bm{k}'}^{\dagger}c_{\bm{k}}^{}
\end{equation}
with the hole number operator
\begin{equation}
	N_{\rm h}=\sum_{\bm{q}}d_{\bm{q}}^{\dagger}d_{\bm{q}}^{}\,.
\end{equation}
Acting on the vacuum state without electrons and holes, both sides in Eq.~\eqref{eq:low_density_e} are identical. For any neutral semiconductor subject to optical driving, electrons and holes are created pairwise, such that acting on the state with one electron present, i.e., one hole present, will yield $N_{\rm h}\rightarrow 1$ and again both sides in Eq.~\eqref{eq:low_density_e} are identical. For larger numbers of electrons and holes the relation breaks down, such that it is viable in the regime of weak optical driving only~\cite{axt1994dynamics, katsch2018theory}. In a similar fashion we can write
\begin{equation}\label{eq:low_density_h}
	d_{\bm{k}'}^{\dagger}d_{\bm{k}}^{}\approx \sum_{\bm{q}}c_{\bm{q}}^{\dagger}d_{\bm{k}'}^{\dagger}d_{\bm{k}}^{}c_{\bm{q}}^{}\,.
\end{equation}
Using Eqs.~\eqref{eq:invert_def_Y_dag}, \eqref{eq:low_density_e} and \eqref{eq:low_density_h}, we can bring the Hamiltonian in Eq.~\eqref{eq:H_eh} into a form where we can apply the Wannier equation~\eqref{eq:wannier} to obtain
\begin{equation}
	H_{\rm eh}\approx \sum_{l,\bm{Q}} E_{l,\bm{Q}}Y_{l,\bm{Q}}^{\dagger}Y_{l,\bm{Q}}^{}\,.
\end{equation}
Restricting ourselves to the lowest-lying 1s exciton, dropping the index $l$, yields the homogeneous exciton Hamiltonian in Eq.~\eqref{eq:ex-hom}.

We can apply the same relations to derive the exciton-pho\-non interaction in Eq.~\eqref{eq:V_ex_ph}, starting from the interaction of electrons and holes with pho\-nons
\begin{align}
	V_{\rm e-p}&=\hbar\sum_{j,\bm{K},\bm{Q}}\qty(g^{({\rm e})}_{j,\bm{Q}}c_{\bm{K}+\bm{Q}}^{\dagger}c_{\bm{K}}^{}-g^{({\rm h})}_{j,\bm{Q}}d_{\bm{K}+\bm{Q}}^{\dagger}d_{\bm{K}}^{})\notag\\
	&\qquad\qquad\times\qty(b_{j,\bm{Q}}^{}+b_{j,-\bm{Q}}^{\dagger})\,.
\end{align}
Using Eqs.~\eqref{eq:invert_def_Y_dag}, \eqref{eq:low_density_e} and \eqref{eq:low_density_h}, we obtain
\begin{align}
	V_{\rm e-p}=&\hbar \sum_{j,\bm{K},\bm{Q}, l, l'}\qty[g_{j,\bm{Q}}^{(\rm e)}F_{l,l'}(\mu_{\rm h}\bm{Q})-g_{j,\bm{Q}}^{(\rm h)}F_{l,l'}(-\mu_{\rm e}\bm{Q})]\notag\\
	&\qquad\times Y_{l',\bm{K}+\bm{Q}}^{\dagger}Y_{l,\bm{K}}^{}\qty(b_{j,\bm{Q}}^{}+b_{j,-\bm{Q}}^{\dagger})\,
\end{align}
with the exciton form factors
\begin{equation}
	F_{l,l'}(\bm{Q})=\sum_{\bm{K}}\Phi_l(\bm{K})^{}\Phi^*_{l'}(\bm{K}+\bm{Q})\,.
\end{equation}
Dropping again the indices $l$ and $l'$, focusing on the 1s exciton, yields the exciton-pho\-non interaction in Eq.~\eqref{eq:V_ex_ph}.

\section{Derivation of the moiré exciton picture Hamiltonian}\label{app:moire_exciton_picture}
The solutions of the moiré exciton eigenvalue equation~\eqref{eq:eigenvalue_moire} form a complete
\begin{equation}
	\sum_n \varphi^{(n)*}_{\bm{k}+\bm{G}'}\varphi^{(n)}_{\bm{k}+\bm{G}}=\delta_{\bm{G}\bm{G}'}\label{eq:moire_complete}
\end{equation}
and orthonormal
\begin{equation}
	\sum_{\bm{G}}\varphi^{(n)*}_{\bm{k}+\bm{G}}\varphi^{(n')}_{\bm{k}+\bm{G}}=\delta_{nn'}\label{eq:moire_ortho}
\end{equation}
set of functions for each $\bm{k}$, since the matrix 
\begin{equation}
	A_{\bm{G}\bm{G}'}(\bm{k})=E_{\bm{k}+\bm{G}'}\delta_{\bm{G},\bm{G}'}+V_{\bm{G}-\bm{G}'}
\end{equation}
appearing in the eigenvalue equation~\eqref{eq:eigenvalue_moire} is hermitian. The moire exciton operators defined in Eq.~\eqref{eq:X_dag} obey bosonic commutation relations in the low-densi\-ty limit, inherited from the approximately bosonic homogeneous exciton operators, which can be shown using Eq.~\eqref{eq:moire_ortho}
\begin{align}
	\comm{X_{n,\bm{k}}^{}}{X_{n',\bm{k}'}^{\dagger}}&=\sum_{\bm{G}\bm{G}'}\varphi^{(n)*}_{\bm{k}+\bm{G}} \varphi^{(n')}_{\bm{k}'+\bm{G}'}\comm{Y_{\bm{k}+\bm{G}}^{}}{Y^{\dagger}_{\bm{k}'+\bm{G}'}}\notag\\
	&=\sum_{\bm{G}\bm{G}'}\varphi^{(n)*}_{\bm{k}+\bm{G}} \varphi^{(n')}_{\bm{k}'+\bm{G}'}\delta_{\bm{k}\bm{k}'}\delta_{\bm{G}\bm{G}'}\notag\\
	&=\delta_{\bm{k}\bm{k}'}\delta_{nn'}\,.
\end{align}

Analog to Eq.~\eqref{eq:invert_def_Y_dag}, we can invert the definition of the moire exciton creation operator in Eq.~\eqref{eq:X_dag} to obtain
\begin{equation}\label{eq:invert_X_dag}
	\sum_n \varphi^{(n)*}_{\bm{k}+\bm{G}}X_{n,\bm{k}}^{\dagger}=Y^{\dagger}_{\bm{k}+\bm{G}}\,.
\end{equation}
Using this relation, we can replace the homogeneous exciton operators in $H_{\rm ex-hom}$ [Eq.~\eqref{eq:ex-hom}] and in $V_{\rm m}$ [Eq.~\eqref{eq:V_m}] to obtain the moire exciton Hamiltonian in Eq.~\eqref{eq:H_ex}. Similarly, Eq.~\eqref{eq:invert_X_dag} is used to derive Eq.~\eqref{eq:V_ex_ph_moire} starting from Eq.~\eqref{eq:V_ex_ph} and Eq.~\eqref{eq:V_ex_laser} starting from Eq.~\eqref{eq:V_ex_laser_original}.

\section{Derivation of the TCL master equation}\label{app:TCL}
In the following we derive the TCL master equation~\eqref{eq:TCL}, applying the Ehrenfest theorem~\cite{breuer2002theory,lengers2020theory, jurgens2024theory}
\begin{equation}
	\dv{t}\expval{A}=-\frac{i}{\hbar}\expval{\comm{A}{H}}
\end{equation}
for operators $A$ that do not depend on time explicitly. The total Hamiltonian
\begin{equation}
	H(t)=H_{\rm ex}+H_{\rm ph}+ V_{\rm ex-ph}+ V_{\rm ex-laser}(t)
\end{equation}
is the sum of the ones given in Eqs.~\eqref{eq:H_ex}, \eqref{eq:H_ph}, \eqref{eq:V_ex_ph_moire} and \eqref{eq:V_ex_laser}. Note that the moire exciton operators, as well as the pho\-non operators appearing in this total Hamiltonian obey bosonic commutation relations.

For the microscopic polarizations $p_{n,\bm{k}}=\expval{X_{n,\bm{k}}}$ we thus obtain (suppressing explicit time dependencies for simplicity)
\begin{align}\label{eq:eom_pol_initial}
	\dv{t}p_{n,\bm{k}}&=-i\omega_{n,\bm{k}}p_{n,\bm{k}}-i\underset{\sigma=\pm}{\sum_{n', j,\bm{Q}}}S^{(n,n',\sigma)}_{j,\bm{k},\bm{Q}}\notag\\
	&\qquad+\frac{i}{\hbar}\mathcal{E}M_n\delta_{\bm{k}\bm{0}}
\end{align}
with the pho\-non-assisted polarizations
\begin{subequations}
	\begin{align}
		S^{(n,n',-)}_{j,\bm{k},\bm{Q}}&= \qty(\mathcal{G}^{(n,n')}_{j,\bm{k},-\bm{Q}})^*\expval{X^{}_{n',\bm{k}-\bm{Q}}b^{}_{j,\bm{Q}}}\,,\\
		S^{(n,n',+)}_{j,\bm{k},\bm{Q}}&= \qty(\mathcal{G}^{(n,n')}_{j,\bm{k},-\bm{Q}})^*\expval{X^{}_{n',\bm{k}-\bm{Q}}b^{\dagger}_{j,-\bm{Q}}}\,.
	\end{align}
\end{subequations}
To obtain a closed set of equations for the microscopic polarizations $p_{n,\bm{k}}$, we need to derive the equations of motion for the pho\-non-assisted polarizations, applying again the Ehrenfest theorem. Since we are interested in linear absorption spectra, i.e., weak optical driving, we neglect all terms with more than one exciton operator, such that
\begin{align}
	\dv{t}&S^{(n,n',-)}_{j,\bm{k},\bm{Q}}=-i\qty(\omega_{n',\bm{k}-\bm{Q}}+\Omega_{j,\bm{Q}})S^{(n,n',-)}_{j,\bm{k},\bm{Q}}\notag\\
	&-i\sum_{n''} \qty(\mathcal{G}^{(n,n')}_{j,\bm{k},-\bm{Q}})^*\mathcal{G}^{(n'',n')}_{j,\bm{k},-\bm{Q}}p_{n'',\bm{k}}\notag\\
	&-i\sum_{n'',j',\bm{Q}'} \qty(\mathcal{G}^{(n,n')}_{j,\bm{k},-\bm{Q}})^*\mathcal{G}^{(n'',n')}_{j',\bm{k}-\bm{Q}-\bm{Q}',\bm{Q}'}\times\notag\\
	&\quad\times\left(\expval{X^{}_{n'',\bm{k}-\bm{Q}-\bm{Q}'}b^{}_{j',\bm{Q}'}b^{}_{j,\bm{Q}}}\right.\notag\\
	&\qquad\left.+\expval{X^{}_{n'',\bm{k}-\bm{Q}-\bm{Q}'}b^{\dagger}_{j',-\bm{Q}'}b_{j,\bm{Q}}^{}}\right) \notag\\
	&+\frac{i}{\hbar} \qty(\mathcal{G}^{(n,n')}_{j,\bm{k},-\bm{Q}})^*\mathcal{E}M_{n'}\expval{b_{j,\bm{Q}}}\delta_{\bm{k}\bm{Q}}\,.
\end{align}
At this point we apply a Born approximation, factorizing the expectation values containing the exciton operator and two pho\-non operators. Considering the pho\-nons to be in a thermal state, we have $\expval{b_{j,\bm{Q}}}=\expval{b_{j',\bm{Q}'}b_{j,\bm{Q}}}=0$ and $\expval{b_{j',-\bm{Q}'}^{\dagger}b_{j,\bm{Q}}}=n_{j,\bm{Q}}\delta_{jj'}\delta_{\bm{Q},-\bm{Q}'}$ with the thermal occupation $n_{j,\bm{Q}}$ from Eq.~\eqref{eq:n_phon}. 

With these approximations, and doing the same derivation also for $S^{(n,n',+)}_{j,\bm{k},\bm{Q}}$, we obtain
\begin{align}\label{eq:eom_phon_assisted}
	\dv{t}S^{(n,n',\sigma)}_{j,\bm{k},\bm{Q}}&=-i\qty(\omega_{n',\bm{k}-\bm{Q}}-\sigma\Omega_{j,-\sigma\bm{Q}})S^{(n,n',\sigma)}_{j,\bm{k},\bm{Q}}\\
	&-i\sum_{n''}\qty(\mathcal{G}^{(n,n')}_{j,\bm{k},-\bm{Q}})^*\mathcal{G}^{(n'',n')}_{j,\bm{k},-\bm{Q}}N_{j,\bm{Q}}^{(-\sigma)}p_{n'',\bm{k}}\notag
\end{align}
with $N_{j,\bm{Q}}^{(\sigma)}$ given in Eq.~\eqref{eq:N_phon}.

In the calculation of the linear absorption spectrum, we consider an ultrafast optical $\delta$-pulse acting at $t=0$, creating microscopic exciton polarizations. Since according to Eq.~\eqref{eq:eom_phon_assisted} the pho\-non-assisted polarization is not explicitly driven by the light pulse, we have directly after optical excitation $S^{(n,n',\sigma)}_{j,\bm{k},\bm{Q}}(t=0^+)=0$. We can then formally solve these equations of motion to obtain for $t>0$
\begin{align}
	S^{(n,n',\sigma)}_{j,\bm{k},\bm{Q}}(t)&=-i\sum_{n''}\qty(\mathcal{G}^{(n,n')}_{j,\bm{k},-\bm{Q}})^*\mathcal{G}^{(n'',n')}_{j,\bm{k},-\bm{Q}}N_{j,\bm{Q}}^{(-\sigma)}\\
	&\times\int\limits_{0^+}^t\dd\tau\, p_{n'',\bm{k}}(\tau)e^{-i\qty(\omega_{n',\bm{k}-\bm{Q}}-\sigma\Omega_{j,-\sigma\bm{Q}})(t-\tau)}\notag\,.
\end{align}
Now we apply the TCL approximation, expressing $p_{n'',\bm{k}}(\tau)$ via $p_{n'',\bm{k}}(t)$ using Eq.~\eqref{eq:eom_pol_initial} in zeroth order with respect to the exciton-pho\-non coupling and with $\mathcal{E}(t')=0$ for $0<\tau\leq t'\leq t$ since we are interested in the dynamics after the ultrashort pulse. Note that the pho\-non-assisted polarizations themselves are already quantities of at least second order in the exciton-pho\-non coupling. We thus arrive at 
\begin{align}
	&S^{(n,n',\sigma)}_{j,\bm{k},\bm{Q}}(t)\approx-i\sum_{n''}\qty(\mathcal{G}^{(n,n')}_{j,\bm{k},-\bm{Q}})^*\mathcal{G}^{(n'',n')}_{j,\bm{k},-\bm{Q}}N_{j,\bm{Q}}^{(-\sigma)}\\
	&\qquad\times p_{n'',\bm{k}}(t)\int\limits_{0^+}^t\dd\tau\, e^{-i\qty(\omega_{n',\bm{k}-\bm{Q}}-\omega_{n'',\bm{k}}-\sigma\Omega_{j,-\sigma\bm{Q}})(t-\tau)}\notag\,.
\end{align}
Inserting this expression for the pho\-non-assisted polarizations into Eq.~\eqref{eq:eom_pol_initial} yields the TCL master equation~\eqref{eq:TCL} for arbitrary moiré exciton momenta $\bm{k}$ without the phenomenological radiative decay, i.e.,
\begin{align}
	\dv{t}p_{n,\bm{k}}(t)&=-i\omega_{n,\bm{k}}p_{n,\bm{k}}(t)+\frac{i}{\hbar}\mathcal{E}(t)M_n\delta_{\bm{k}\bm{0}}\notag\\
	&-\sum_{n'',j}\Gamma_{j,\bm{k}}^{(n,n'')}(t)p_{n'',\bm{k}}(t)\,.\label{eq:TCL_k}
\end{align}
All effects due to the thermal pho\-non bath, i.e., pho\-non-in\-duced dissipation and energy renormalization (polaron shifts), are captured in the time dependent dissipation coefficient matrix 
\begin{equation}\label{eq:Gamma_t_k}
	\Gamma_{j,\bm{k}}^{(n,n'')}(t)=\int\dd\Omega\, \rho_{j,\bm{k}}^{(n,n'')}(\Omega)\int\limits_0^t\dd\tau\, e^{-i\Omega\tau}
\end{equation}
due to pho\-nons in branch $j$ with the gPSD given by
\begin{align}\label{eq:g_PSD_k}
	\rho_{j,\bm{k}}^{(n,n'')}(\Omega)&=\sum_{n',\bm{Q},\sigma=\pm}\qty(\mathcal{G}_{j,\bm{k},-\bm{Q}}^{(n,n')})^*\mathcal{G}_{j,\bm{k},\bm{-Q}}^{(n'',n')}N_{j,\bm{Q}}^{(-\sigma)}\\
	&\quad\times\delta\qty(\Omega+\sigma\Omega_{j,-\sigma\bm{Q}}+\omega_{n'',\bm{k}}-\omega_{n',\bm{k}-\bm{Q}})\notag\,.
\end{align}
In the main text we drop the index $\bm{k}$, focusing only on the optically active microscopic polarizations with $\bm{k}=\bm{0}$.
\section{Markov\-ian and non-Markov\-ian dynamics}\label{app:markov_non_markov}
If we discard inter-polarization coupling in the TCL master equation~\eqref{eq:TCL}, i.e., neglect the sum over $n''\neq n$, the equations for the different polarizations decouple. We can separate the polarization dynamics $p_n(t)$ in Eq.~\eqref{eq:TCL} after optical excitation, i.e., for $\mathcal{E}=0$, into a Markov\-ian contribution obeying
\begin{equation}
	\dv{t} p_n^{\rm M}(t)=\qty(-i\omega_{n}-\sum_j \overline{\Gamma}^{(n,n)}_{j}-\frac{\gamma_n}{2})p_n^{\rm M}(t)
\end{equation}
and a non-Markov\-ian contribution obeying
\begin{equation}\label{eq:TCL_nM}
	\dv{t}\Phi_n(t)=\sum_j\qty[\overline{\Gamma}^{(n,n)}_{j}-\Gamma^{(n,n)}_{j}(t)]\Phi_n(t)\,,
\end{equation}
such that $p_n(t)=p_n^{\rm M}(t) \Phi_n(t)$ solves the TCL master equation~\eqref{eq:TCL} given the restrictions stated above. In the long-time limit the dissipation coefficients converge towards the Markov limit in Eq.~\eqref{eq:markov}, such that $\dv{t} \Phi_n\rightarrow 0$ and therefore
\begin{subequations}
\begin{align}
	p_n(t)&\underset{t\rightarrow\infty}{\sim}p_n^{\rm M}(t)=e^{-i\tilde{\omega}_n t-\frac{\tilde{\gamma}_n}{2}t}p_n(0)\,,\\
	\tilde{\omega}_n&=\omega_{n}+\sum_j \Im\qty(\overline{\Gamma}^{(n,n)}_{j})\,,\\
	\tilde{\gamma}_n&=\gamma_n+2\sum_j \Re\qty(\overline{\Gamma}^{(n,n)}_{j})\,.
\end{align}
\end{subequations}
Here we choose the initial conditions $p_n^{\rm M}(0)=p_n(0)$, $\Phi_n(0)=1$. On short time scales when the time dependent dissipation coefficients do not coincide with their Markov limit we can however have $\dv{t}\Phi_n(t)\neq 0$. Given the initial condition from above, we can formally solve Eq.~\eqref{eq:TCL_nM} yielding
\begin{equation}
	\Phi_n(t)=e^{-\phi_n(t)}
\end{equation}
with
\begin{align}
	\phi_n(t)&=\int\limits_0^{t}\dd t'\, \sum_j\qty[\Gamma_{j}^{(n,n)}(t')-\overline{\Gamma}^{(n,n)}_{j}]\\
	&=\int\dd\Omega\, \sum_j\rho_{j}^{(n,n)}(\Omega)\int\limits_0^t\dd t'\,\int\limits_{0}^{t'}\dd\tau e^{-i\Omega \tau}\notag\\
	&\qquad\qquad-\overline{\Gamma}^{(n,n)}_{j}t\notag\,.
\end{align}
To make further progress, especially in evaluating the Markov limit $t\rightarrow\infty$, we introduce a small damping  $\Omega\rightarrow \Omega-i\eta$ and take the limit $\eta\rightarrow 0^+$ at the end of calculations to apply the Dirac identity. Physically such a damping might originate from the decay of the pho\-non-assisted polarizations. With this small damping, we can evaluate the $t$- and $t'$-integral to
\begin{align}
	&\lim\limits_{\eta\rightarrow 0^+}\int\limits_0^t\!\dd t'\!\int\limits_{0}^{t'}\!\dd\tau e^{-i\Omega \tau-\eta \tau}=\lim\limits_{\eta\rightarrow 0^+}\int\limits_0^t\!\dd t' \frac{e^{-i\Omega t'-\eta t'}-1}{-i\Omega -\eta}\notag\\
	&\quad=-i\mathcal{P}\frac{1}{\Omega} t + \pi \delta(\Omega) t+\lim\limits_{\eta\rightarrow 0^+}\int\limits_0^t\!\dd t' \frac{e^{-i\Omega t'-\eta t'}}{-i\Omega -\eta}\,.
\end{align}
Here, $\mathcal{P}$ denotes the principal value and we used the Dirac identity~\cite{breuer2002theory}. With this intermediate result, using the dissipation coefficients in the Markov limit from Eq.~\eqref{eq:markov}, we can write
\begin{align}
	\phi_n(t)&=\lim\limits_{\eta\rightarrow 0^+}\int\dd\Omega\, \sum_j\rho_{j}^{(n,n)}(\Omega)\int\limits_0^t\dd t'\, \frac{e^{-i\Omega t'-\eta t'}}{-i\Omega -\eta}\notag\\
	&=\lim\limits_{\eta\rightarrow 0^+}\int\dd\Omega\, \sum_j\rho_{j}^{(n,n)}(\Omega)\frac{e^{-i\Omega t-\eta t}-1}{\qty(-i\Omega -\eta)^2}\,.\label{eq:phi_n}
\end{align}

\section{Separation of spectra into ZPL and PSBs}\label{app:zpl_psbs}
Discarding inter-polarization coupling, as discussed in App.~\ref{app:markov_non_markov}, the solutions $p_{n}(t)$ to the TCL master equation~\eqref{eq:TCL} after optical excitation, i.e., for $\mathcal{E}=0$, are given by
\begin{subequations}\label{eq:pol_n}
	\begin{align}
		p_n(t)&=e^{-i\tilde{\omega}_{n}t-\frac{\tilde{\gamma}_n}{2} t}e^{-\phi_n(t)}p_{n}(0)\,,\\
		\tilde{\omega}_n&=\omega_{n}+\sum_j \Im\qty(\overline{\Gamma}^{(n,n)}_{j})\,,\\
		\tilde{\gamma}_n&=\gamma_n+2\sum_j \Re\qty(\overline{\Gamma}^{(n,n)}_{j})\,,\\
		\phi_n(t)&=\lim\limits_{\eta\rightarrow 0^+}\int\dd\Omega\, \sum_j\frac{\rho_{j}^{(n,n)}(\Omega)}{\qty(\Omega-i\eta)^2}\notag\\
		&\qquad\times\qty[1-e^{-i(\Omega-i\eta) t}]\,,\label{eq:phi_n_summary}
	\end{align}
\end{subequations}
where the limit $\eta\rightarrow 0^+$ is taken at the end of calculations. Here, $\tilde{\omega}_n$ and $\tilde{\gamma}_n$ contain the polaron shift and the pho\-non-in\-duced dissipation in the Markov limit, respectively [see Eq.~\eqref{eq:markov}]. The dephasing function $\phi_n(t)$ describes the impact of non-Markov\-ian dynamics on the optically active moiré exciton polarization.

The polarization dynamics in Eqs.~\eqref{eq:pol_n} can be used to gain analytical insight into the absorption spectrum in Eq.~\eqref{eq:alpha_omega} by performing a power series in $\phi_n(t)$ and using the initial condition given in Eq.~\eqref{eq:initia_p}, such that
\begin{equation}
	\alpha(\omega)\sim \sum_{n,q}\alpha_{n}^{(q)}(\omega)
\end{equation}
with
\begin{equation}
	\alpha_{n}^{(q)}(\omega)=|M_n|^2 \Re\qty{\int\limits_0^{\infty}\dd t\,e^{i(\omega-\tilde{\omega}_n) t-\frac{\tilde{\gamma}_n}{2}t}\frac{\qty[-\phi_n(t)]^q}{q!}}\,.
\end{equation}
The zeroth-order contribution with respect to the non-Markov\-ian dephasing $\phi_n(t)$, i.e., the ZPL, is given by
\begin{equation}
	\alpha_{n}^{(0)}(\omega)= |M_n|^2\frac{\tilde{\gamma}_n/2}{\qty(\tilde{\gamma}_n/2)^2+(\omega-\tilde{\omega}_{n})^2}\,.
\end{equation}
The first-order contribution to the PSBs reads
\begin{align}
	&\alpha_{n}^{(1)}(\omega)=-|M_n|^2 \Re\qty[\int\limits_0^{\infty}\!\dd t\,e^{i(\omega-\tilde{\omega}_n) t-\frac{\tilde{\gamma}_n}{2}t}\phi_n(t)]\\
	&=\lim\limits_{\eta\rightarrow 0^+}|M_n|^2\Re\left\lbrace\int\dd\Omega\, \sum_j\frac{\rho_{j}^{(n,n)}(\Omega)}{\qty(\Omega-i\eta)^2}\right.\notag\\
	&\left.\times\qty[\frac{1}{\frac{\tilde{\gamma}_n}{2}+\eta-i(\omega-\Omega-\tilde{\omega}_n)}-\frac{1}{\frac{\tilde{\gamma}_n}{2}-i(\omega-\tilde{\omega}_n)}]\right\rbrace\notag\\
	&=\mathcal{P}\int\dd\Omega\, \sum_j\frac{\rho_{j}^{(n,n)}(\Omega)}{\Omega^2}\qty[\alpha_{n}^{(0)}(\omega-\Omega)-\alpha_{n}^{(0)}(\omega)]\notag\\
	&\quad +\pi \int\dd\Omega\, \sum_j\rho_{j}^{(n,n)}(\Omega)\delta^{(1)}(\Omega)\notag\\
	&\quad\times\qty[\frac{\omega-\Omega-\tilde{\omega}_n}{\tilde{\gamma}_n/2}\alpha_{n}^{(0)}(\omega-\Omega)-\frac{\omega-\tilde{\omega}_n}{\tilde{\gamma}_n/2}\alpha_{n}^{(0)}(\omega)]\notag\\
	&=\mathcal{P}\int\dd\Omega\, \sum_j\frac{\rho_{j}^{(n,n)}(\Omega)}{\Omega^2}\qty[\alpha_{n}^{(0)}(\omega-\Omega)-\alpha_{n}^{(0)}(\omega)]\notag\\
	&+\pi  \sum_j\frac{\rho_{j}^{(n,n)}(0)}{\tilde{\gamma}_n/2}\qty[\alpha_{n}^{(0)}(\omega)+(\omega-\tilde{\omega}_n)\alpha_{n}^{(0)'}(\omega)]\notag\,.
\end{align}
Here, $\delta^{(1)}(\Omega)$ denotes the first derivative of the $\delta$-function and we made use of the Dirac identity, partial integration and the fact that the gPSD $\rho_j^{(n,n)}$ is real.

	
	\section{$C_{3v}$ point group symmetry of the moiré exciton eigenvalue equation}\label{app:symmetries}
	The moiré eigenvalue equation~\eqref{eq:eigenvalue_moire} obeys some important symmetries stemming from the symmetries of the moiré exciton potential $V_{\bm{G}}$ in Eq.~\eqref{eq:V_G}. These are especially useful for the classification of solutions at $\bm{k}=\bm{0}$, i.e., for the potentially bright moiré excitons. For this special value Eq.~\eqref{eq:eigenvalue_moire} becomes
	\begin{equation}\label{eq:eigenvalue_moire_k0}
		\sum_{\bm{G}'}\qty(E_{\bm{G}'}\delta_{\bm{G}\bm{G}'}+V_{\bm{G}-\bm{G}'})\varphi_{\bm{G}'}^{(n)}=\hbar\omega_{n}\varphi_{\bm{G}}^{(n)}
	\end{equation}
	with $\omega_n=\omega_{n,\bm{0}}$. The following symmetry considerations rely on the fact that the homogeneous exciton energy only depends on the absolute value of the momentum $E_{\bm{K}}=E_{|\bm{K}|}$. For any orthogonal transformation $\underline{S}$ mapping wavevectors according to $\bm{K}\rightarrow \underline{S}\bm{K}$ while retaining their length $|\underline{S}\bm{K}|=|\bm{K}|$, which additionally obeys $V_{\underline{S}\bm{G}}=V_{\bm{G}}$, i.e., is a symmetry of the moiré exciton potential, we find that $\varphi_{\underline{S}\bm{G}}^{(n)}$ is an eigenfunction of Eq.~\eqref{eq:eigenvalue_moire_k0} with the same eigenvalue $\hbar\omega_n$ as $\varphi_{\bm{G}}^{(n)}$. This allows us to classify the solutions $\varphi_{\bm{G}}^{(n)}$ via their properties with regards to the symmetry transformations $\underline{S}$ of the moire exciton potential. 
	
	The potential in Eq.~\eqref{eq:V_G} is invariant under rotation by $120^\circ$, as well as reflection along the three axes running trough $\bm{G}_j$ and $\bm{G}_{[(j+2)\mod 6]+1}$ with $j=1,3,5$. The corresponding symmetry group is the $C_{3v}$ point group~\cite{li1985solution, ruiz2020theory}. Defining the polar angle $\phi$ in two dimensions relative to the axis running through $\bm{G}_1$ and $\bm{G}_4$, i.e., $\bm{G}_1$ points in positive $x$-direction, the action of a $120^\circ$ rotation is given by $U(2\pi/3) f(\phi)\equiv\bar{U}f(\phi)=f(\phi+2\pi/3)$.
	
	Acting on a Hilbert space, such rotations are represented by unitary operators and their form can be deduced by considering infinitesimal rotations
	\begin{equation}
		U(\epsilon)f(\phi)=f(\phi+\epsilon)=f(\phi)+\epsilon\pdv{\phi}f(\phi)+\mathcal{O}(\epsilon^2)\,
	\end{equation}
	implying
	\begin{equation}
		U(\epsilon)=1+\epsilon\pdv{\phi}+\mathcal{O}(\epsilon^2)\,.
	\end{equation}
	Any finite rotation can be constructed via
	\begin{align}
		U(\theta)&=\lim\limits_{N\rightarrow\infty}U(\theta/N)^N=\lim\limits_{N\rightarrow\infty}\qty(1+\frac{\theta}{N}\pdv{\phi})^N\notag\\
		&=e^{\theta \pdv{\phi}}\,.
	\end{align}
	The eigenstates of $\bar{U}=U(2\pi/3)$ are given by the complete set of periodic functions $e^{im\phi}$ with $m\in\mathbb{Z}$ and
	\begin{equation}
		\bar{U}e^{im\phi}=e^{im\frac{2\pi}{3}}e^{im\phi}\,.
	\end{equation}
	Due to the rotation being discrete we cannot distinguish between eigenfunctions that differ in the quantum number $m$ by multiples of $3$, since $e^{i(m+3)\frac{2\pi}{3}}=e^{im\frac{2\pi}{3}}$. We thus get three classes of eigenfunctions $f_0(\phi)$ and $f_\pm(\phi)$ with
	\begin{subequations}
		\begin{align}
			\bar{U}f_0(\phi)&=f_0(\phi+2\pi/3)=f_0(\phi)\,,\\
			\bar{U}f_{\pm}(\phi)&=f_{\pm}(\phi+2\pi/3)=e^{\pm i\frac{2\pi}{3}}f_{\pm}(\phi)\,.
		\end{align}
	\end{subequations}
	These eigenfunctions have the respective quantum numbers $\bar{m}=0$ and $\bar{m}=\pm 1$ with $\bar{m}=[(m+1)\mod 3]-1$ and are superpositions of the corresponding functions $e^{im\phi}$ with the same value of $\bar{m}$.
	
	The reflection along the axis running through $\bm{G}_1$ and $\bm{G}_4$, called $\sigma_v$ in the following, flips the angle $\sigma_v f(\phi)=f(-\phi)$. Acting on the rotation eigenfunctions with the reflection operation yields
	\begin{align}
		\sigma_v \bar{U} f_0(\phi)&=\sigma_v f_0(\phi)=f_0(-\phi)=\bar{U}f_0(-\phi)\notag\\&=\bar{U}\sigma_v f_0(\phi)\,,
	\end{align}
	as well as
	\begin{subequations}
		\begin{align}
			\sigma_v \bar{U} f_{\pm}(\phi)&=e^{\pm i \frac{2\pi}{3}} \sigma_v f_{\pm}(\phi)=e^{\pm i \frac{2\pi}{3}} f_{\pm}(-\phi)\,,\\
			\bar{U}\sigma_v f_{\pm}(\phi)&=\bar{U} f_{\pm}(-\phi)=f_{\pm}(-\phi-2\pi/3)\notag\\&=e^{\mp i\frac{2\pi}{3}} f_{\pm}(-\phi)\,.
		\end{align}
	\end{subequations}
	On the subspace of rotational eigenfunctions with quantum number $\bar{m}=0$ the reflection and rotation obviously commute and we can characterize the eigenfunction via their transformation property under reflection. Since two reflections along the same axis do nothing, i.e., $\sigma_v^2=1$, $\sigma_v$ has the eigenvalues $\pm 1$ and we find common eigenstates $f_0^{\pm}(\phi)$ with
	\begin{subequations}
	\begin{align}
		\bar{U} f_0^{(\pm)}(\phi)&=f_0^{(\pm)}(\phi)\,,\\
		\sigma_v f_0^{(\pm)}(\phi)&=\pm f_0^{(\pm)}(\phi)\,.
	\end{align}
	\end{subequations}
	These belong to the one-dimensional irreps $A_1$ and $A_2$ of even $(+)$ and odd $(-)$ functions under reflection and invariance under three-fold rotation of the $C_{3v}$ point group.
	
	On the subspace of rotational eigenfunctions with quantum number $\bar{m}=\pm 1$ on the other hand the operators $\sigma_v$ and $\bar{U}$ do not commute. In fact we find that $\sigma_v$ connects the subspaces with $\bar{m}=\pm 1$ since
	\begin{equation}
		\bar{U}\sigma_vf_{\pm}(\phi)=e^{\mp i\frac{2\pi}{3}}\sigma_v f_{\pm}(\phi)\,,
	\end{equation}
	i.e., $\sigma_v f_{\pm}(\phi)$ behaves as $f_{\mp}(\phi)$ under the rotation $\bar{U}$. The eigenfunctions $f_{\pm}(\phi)$ belong to the two-dimensional irrep $E$ of the point group $C_{3v}$.
	
	With these considerations, the eigenfunctions $\varphi_{\bm{G}}^{(n)}$ either belong to the one-dimensional irreps $A_{1/2}$ obeying $\bar{U}\varphi_{\bm{G}}^{(n)}=\varphi_{\underline{R}\bm{G}}^{(n)}=\varphi_{\bm{G}}^{(n)}$ or to the two-dimensional irrep $E$ obeying $\bar{U}\varphi_{\bm{G}}^{(n)}=\varphi_{\underline{R}\bm{G}}^{(n)}=e^{\pm i\frac{2\pi}{3}}\varphi_{\bm{G}}^{(n)}$. Here, $\underline{R}$ is the orthogonal rotation matrix describing a $120^\circ$ rotation. This implies that the solutions belonging to the irrep $E$ describe dark moiré excitons since they have to obey $\varphi_{\bm{0}}^{(n)}=\varphi_{\underline{R}\bm{0}}^{(n)}=e^{\pm i\frac{2\pi}{3}}\varphi_{\bm{0}}^{(n)}=0$ leading to a vanishing dipole matrix element $M_n$ and thus no coupling to light in Eq.~\eqref{eq:V_ex_laser}. With the same argument the solutions belonging to the irrep $A_2$ which is antisymmetric under reflection, are also dark since they obey $\varphi_{\bm{0}}^{(n)}=-\varphi_{\bm{0}}^{(n)}$. 
	
	\section{Comments on the inter-polari\-za\-tion coupling}\label{app:inter_pol}
	We found numerically that the inter-polarization coupling in the master equation~\eqref{eq:TCL}, i.e., terms with $n''\neq n$, are typically negligible. In the following we will discuss some analytical properties of the theory that support this finding. Since there is no general proof of the smallness of these terms, there could still be situations where this coupling becomes important, which we however did not encounter so far.
	\subsection{Short time scale behavior of the dissipation coefficients}\label{app:gPSD_props}
	Integrating the gPSD in Eq.~\eqref{eq:g_PSD_k} which is the generalization of Eq.~\eqref{eq:g_PSD} to arbitrary $\bm{k}$ and using Eqs.~\eqref{eq:g_moire} and \eqref{eq:formfactor_moire} yields
	\begin{align}\label{eq:g_PSD_int_1}
		\int\dd\Omega\,&\rho_{j,\bm{k}}^{(n,n'')}(\Omega)=\sum_{n',\bm{Q},\sigma=\pm}\qty(\mathcal{G}_{j,\bm{k},-\bm{Q}}^{(n,n')})^*\mathcal{G}_{j,\bm{k},\bm{-Q}}^{(n'',n')}N_{j,\bm{Q}}^{-\sigma}\notag\\
		&=\sum_{n',\bm{G},\bm{G}',\bm{Q},\sigma}|g_{j,-\bm{Q}}|^2 \varphi^{(n')}_{\bm{k}-\bm{Q}+\bm{G}}\qty(\varphi_{\bm{k}+\bm{G}}^{(n)})^*\notag\\
		&\qquad\qquad\times \qty(\varphi^{(n')}_{\bm{k}-\bm{Q}+\bm{G}'})^*\varphi_{\bm{k}+\bm{G}'}^{(n'')}N_{j,\bm{Q}}^{-\sigma}\,.
	\end{align}
	The completeness of the $\varphi_{\bm{k}+\bm{G}}^{(n)}$ [Eq.~\eqref{eq:moire_complete}] together with the sum over $n'$ yields a $\delta_{\bm{G}\bm{G}'}$, such that
	\begin{align}\label{eq:g_PSD_int_2}
		\int\dd\Omega\,\rho_{j,\bm{k}}^{(n,n'')}(\Omega)&=\sum_{\bm{G},\bm{Q},\sigma}|g_{j,-\bm{Q}}|^2 \qty(\varphi_{\bm{k}+\bm{G}}^{(n)})^*\notag\\
		&\qquad\times\varphi_{\bm{k}+\bm{G}}^{(n'')}N_{j,\bm{Q}}^{-\sigma}\,.
	\end{align}
	Orthonormality of the $\varphi_{\bm{k}+\bm{G}}^{(n)}$ [Eq.~\eqref{eq:moire_ortho}] together with the sum over $\bm{G}$ then leads to
	\begin{align}\label{eq:g_PSD_int_3}
		\int\dd\Omega\,\rho_{j,\bm{k}}^{(n,n'')}(\Omega)&=\delta_{n,n''}\sum_{\bm{Q},\sigma}|g_{j,-\bm{Q}}|^2 N_{j,\bm{Q}}^{-\sigma}\,.
	\end{align}
	On very short time scales where $t\ll\overline{\Omega}_{(n,n'')}^{-1}$ with $\overline{\Omega}_{(n,n'')}$ being typical frequencies contained in  $\rho_{j}^{(n,n'')}$, this implies that the pho\-non-in\-duced transitions described by the rates in Eq.~\eqref{eq:Gamma_t} obey
	\begin{equation}
		\Gamma_{j}^{(n,n'')}\qty(t\ll\overline{\Omega}_{(n,n'')}^{-1})\sim \delta_{n,n''}\,,
	\end{equation}
	i.e., there is no coupling between microscopic polarizations from different bands in the TCL equation~\eqref{eq:TCL} on time scales $t\ll\min[\overline{\Omega}_{(n,n'')}^{-1}]$.
	
	\subsection{Long-wavelength behavior of the moiré exciton form factor}\label{app:long_wavelength}
	The moiré exciton form factor for excitons that can in principal be excited optically, i.e., with $\bm{k}=\bm{0}$, reads [see Eq.~\eqref{eq:formfactor_moire}]
	\begin{equation}\label{eq:formfactor_moire_k=0}
		f_{\bm{0}\bm{Q}}^{(n,n')}=\sum_{\bm{G}}\qty(\varphi^{(n')}_{\bm{Q}+\bm{G}})^*\varphi_{\bm{G}}^{(n)}\,.
	\end{equation}
	At $\bm{Q}=\bm{0}$ we can make use of the orthonormality of the $\varphi_{\bm{G}}^{(n)}$ from Eq.~\eqref{eq:moire_ortho}, such that
	\begin{equation}\label{eq:moire_FF_delta}
		f_{\bm{0}\bm{0}}^{(n,n')}=\sum_{\bm{G}}\qty(\varphi^{(n')}_{\bm{G}})^*\varphi_{\bm{G}}^{(n)}=\delta_{nn'}\,.
	\end{equation}
	This implies that the contribution of long-wavelength pho\-nons with $\bm{Q}\rightarrow\bm{0}$ to the gPSD in Eq.~\eqref{eq:g_PSD} is negligible for $n''\neq n$. 
	
	For arbitrary $\bm{Q}$ the moiré exciton form factors fulfill the sum rule
	\begin{align}
		\sum_{n'}\qty|f_{\bm{0}\bm{Q}}^{(n,n')}|^2&=\sum_{n'\bm{G}\bm{G}'}\qty(\varphi^{(n')}_{\bm{Q}+\bm{G}})^*\varphi_{\bm{G}}^{(n)}\varphi^{(n')}_{\bm{Q}+\bm{G}'}\qty(\varphi_{\bm{G}'}^{(n)})^*\notag\\
		&=\sum_{\bm{G}\bm{G}'}\delta_{\bm{G}\bm{G}'}\varphi_{\bm{G}}^{(n)}\qty(\varphi_{\bm{G}'}^{(n)})^*=1\,,
	\end{align}
	where we again made use of the completeness and orthonormality discussed in App.~\ref{app:moire_exciton_picture}. This implies that any growth of the off-diagonals (inter\-band form factors) $n\neq n'$ in Eq.~\eqref{eq:formfactor_moire_k=0} has to stem from a decrease in the diagonal (intra\-band form factors) $n=n'$ for any chosen band $n$. According to Eq.~\eqref{eq:moire_FF_delta} the intra\-band form factors have unit value at $\bm{Q}=\bm{0}$, forcing the inter\-band form factors to vanish there.

	\begin{figure}
		\centering
		\includegraphics[width=0.7\linewidth]{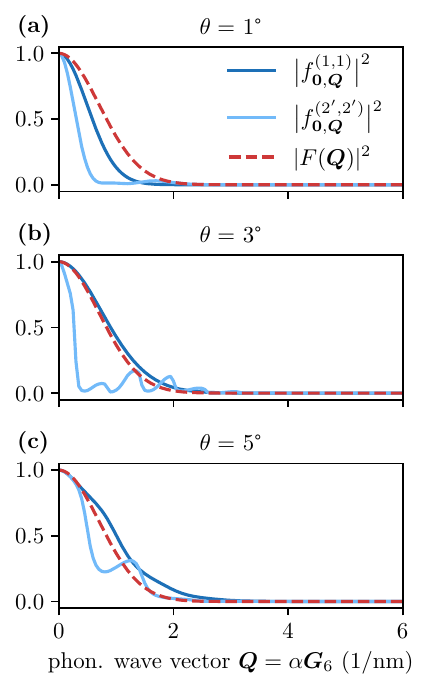}
		\caption{Absolute square of the intra\-band moiré exciton form factors $f_{\bm{0},\bm{Q}}^{(n,n)}$ along a high-symmetry direction for the two lowest lying bright bands $n=1$, $n=2'$ and different twist angles as indicated in the titles (blue). In addition each figure shows the absolute square of the homogeneous exciton form factor from Eq.~\eqref{eq:exc_FF} (red). Here, $n=2'$ indicates the second lowest lying bright moiré exciton band. In terms of the energetic order of the bands, $n=2'$ means $n=4$ for $\theta=1^{\circ}$ and $\theta=3^{\circ}$, as well as $n=2$ for $\theta=5^{\circ}$ (see Fig.~\ref{fig:dipole_MBZtutorial}).}
		\label{fig:fformfactorall}
	\end{figure}
	
	For the smallest twist angle of $\theta=1^{\circ}$ considered in this work, in Fig.~\ref{fig:fformfactorall}~(a) we show the absolute square of the intra\-band moiré exciton form factors $|f^{(n,n)}_{\bm{0},\bm{Q}}|^2$ for the two lowest lying bright moiré exciton bands, i.e., those bands with non-vanishing dipole matrix element at the $\gamma$-point (see Fig.~\ref{fig:dipole_MBZtutorial}), along a high-symmetry direction (blue). We see that they decay from unity at $\bm{Q}=\bm{0}$ on a shorter $\bm{Q}$-scale than the red curve which is the absolute square of the homogeneous exciton form factor from Eq.~\eqref{eq:exc_FF}. However for $|\bm{Q}|<0.5$~nm$^{-1}$ they are close to unity, implying that the corresponding inter\-band form factors $f^{(n,n'\neq n)}_{\bm{0},\bm{Q}}$ are negligible in this range of wavevectors. When we increase the twist angle, the range of $\bm{Q}$ values where inter\-band form factors are negligible typically increases, as can be seen in Fig.~\ref{fig:fformfactorall}~(b, c), which shows the same situation as Fig.~\ref{fig:fformfactorall}~(a), but for $\theta=3^{\circ}$ and $\theta=5^{\circ}$, respectively. 
	
	This discussion shows that for a considerable range of relevant $\bm{Q}$-values, which are determined by the homogeneous form factor, inter\-band moiré exciton form factors are negligible. In this range of $\bm{Q}$-values around the long-wavelength limit $\bm{Q}\rightarrow\bm{0}$ there are no inter\-band contributions with $n\neq n'$ or $n'\neq n''$ to the gPSD in Eq.~\eqref{eq:g_PSD}. If this range of $\bm{Q}$-values dominates the properties of the gPSD, the corresponding dissipation coefficients in Eq.~\eqref{eq:Gamma_t} approximately vanish for $n\neq n''$ leading to a negligible inter-polarization coupling in Eq.~\eqref{eq:TCL}.
	\subsection{Markov limit of the dissipation coefficients}
	The Markov limit decay rates in Eq.~\eqref{eq:markov} are determined by energy-conserving transitions, i.e., $\Omega=0$ in the gPSD in Eq.~\eqref{eq:g_PSD}. The TCL master equation~\eqref{eq:TCL} in the Markov limit and in the frame rotating with the moiré exciton frequencies reads
	\begin{subequations}
		\begin{align}
			\dv{t}\tilde{p}_n(t)&=-\sum_{n'',j}\overline{\Gamma}_j^{(n,n'')}e^{i(\omega_n-\omega_{n''})t}\tilde{p}_{n''}(t)\,,\\
			\tilde{p}_n(t)&=p_n(t) e^{i\omega_n t}\,,
		\end{align}
	\end{subequations}
	where we neglected radiative decay for simplicity and consider the moiré exciton dynamics after optical excitation. We found typical Markov limit decay rates for $n=n''$ to be on the order of $\sim1$~ps$^{-1}$, sometimes going as high as $\sim10$~ps$^{-1}$ at elevated temperatures and special twist angles (see Figs.~\ref{fig:dynamics_ac_4K} and \ref{fig:dynamics_ac_70K_200K}). The decay rates for $n\neq n''$, whose impact we want to discuss in the following, were typically smaller due to the long-wavelength properties discussed in the previous App.~\ref{app:long_wavelength}.

	For the following argument we assume Markov limit dissipation coefficients $\sum_j \overline{\Gamma}_j^{(n,n'')}$ with an absolute value on the order of $1$~ps$^{-1}$ for $n\neq n''$, i.e., for inter-polarization coupling. We can perform a rotating wave approximation for the off-diagonal terms in the previous equation and neglect any terms with $\hbar|\omega_n-\omega_{n''}|\gg \hbar|\sum_j \overline{\Gamma}_j^{(n,n'')}|\sim 1~$meV. Looking at the $\gamma$-point of the band structure in Fig.~\ref{fig:band_structure}, the bands are often sufficiently far away from each other to guarantee this, except for the cases where the bands are exactly degenerate at the $\gamma$-point due to the $C_{3v}$ symmetry, discussed in App.~\ref{app:symmetries}. 
	
	Exact degeneracy at the $\gamma$-point implies that the energy conserving transitions in Eq.~\eqref{eq:g_PSD} due to acoustic pho\-nons are via long-wavelength modes, such that these bands cannot couple via acoustic pho\-nons (see previous App.~\ref{app:long_wavelength}). For optical pho\-nons there can only be energy-conserving transitions between bands that are degenerate at the $\gamma$-point, if one of the bandwidths surpasses the finite optical pho\-non energy $\hbar\Omega_{\rm{opt}}$.
	
	From this discussion on the Markov limit dynamics we tentatively deduce that the rotating wave approximation together with the long-wavelength behavior of the moiré exciton form factors discussed in the previous App.~\ref{app:long_wavelength} seem to constitute the key ingredients that are required to understand why the inter-polarization coupling has negligible influence in the TCL master equation~\eqref{eq:TCL}.
	
	\section{Polarization dynamics and time dependent decay rates due to acoustic pho\-nons at elevated temperatures}\label{app:pol_dyn_T_high}
	
	\begin{figure}[h]
		\centering
		\includegraphics[width=\linewidth]{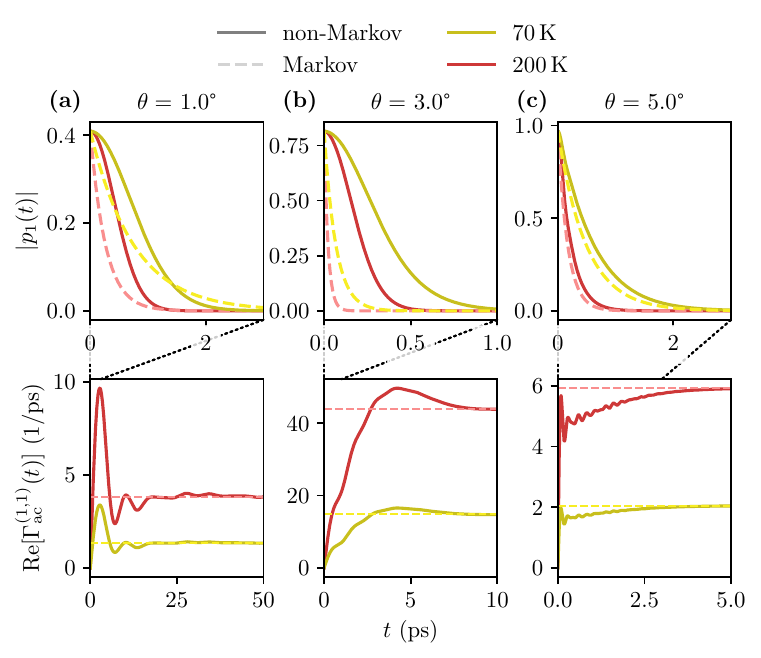}
		\caption{Dynamics of the absolute value of the moiré exciton polarization $|p_1(t)|$ from Eqs.~\eqref{eq:pol_1} including only acoustic phonon scattering $j=\rm{ac}$ (top) and corresponding time dependent decay rates $\Re\big[\Gamma_{\rm{ac}}^{(1,1)}(t)\big]$ calculated via Eq.~\eqref{eq:Gamma_t} (bottom) for the twist angles $\theta=1^{\circ}$ (a), $\theta=3^{\circ}$ (b), and $\theta=5^{\circ}$ (c) at the temperatures $T=70$~K (yellow) and $T=200$~K (red). We compare the full simulations (solid) with the Markov limit (dashed).}
		\label{fig:dynamics_ac_70K_200K}
	\end{figure}

Figure~\ref{fig:dynamics_ac_70K_200K} shows the dynamics of the absolute value of the moiré exciton polarization $|p_1(t)|$ from Eqs.~\eqref{eq:pol_1} (top). We include here only the impact due to acoustic pho\-nons with $j=\rm{ac}$. The corresponding time dependent decay rate $\Re\big[\Gamma_{\rm{ac}}^{(1,1)}(t)\big]$, calculated via Eq.~\eqref{eq:Gamma_t}, is shown in the bottom row. In addition to the full simulations (solid), we show the results in the Markov limit (dashed), obtained by setting $\phi_1=0$ in the dynamics for $p_1$ from Eqs.~\eqref{eq:pol_1} (top) and replacing the full time dependent decay rate by $\Re\big(\overline{\Gamma}^{(1,1)}_{\rm{ac}}\big)$ (bottom), respectively. The results in Fig.~\ref{fig:dynamics_ac_70K_200K} are obtained for $T=70$~K (yellow) and $T=200$~K (red).

Analogous to the discussion of Fig.~\ref{fig:dynamics_ac_4K} for $T=4$~K in the main text, we find that the polarization dynamics exhibit a strong non-Markovian influence for small twist angles ($\theta=1^{\circ}$, a) and close to the magic angle ($\theta=3^{\circ}$, b). Since phonon absorption processes are enhanced at these elevated temperatures, the Markov rate close to the magic angle for phonon absorption is particularly large (b, bottom). At larger twist angles ($\theta=5^{\circ}$, c), the system reaches the Markov limit on a much shorter time scale (bottom) and the polarization decay (top) is much better described by the Markov limit (dashed), compared with the cases of smaller twist angles (a,b).

\medskip
\noindent\textbf{Acknowledgements} \par 
\noindent The authors acknowledge financial support by the Alexander von Humboldt foundation (AvH) within the research group linkage programme funded by the German Federal Ministry
of Education and Research (BMBF).

\medskip
\noindent\textbf{Conflict of Interest} \par 
\noindent The authors declare no conflict of interest.

\medskip
\noindent\textbf{Data Availability Statement} \par 
\noindent The data that support the findings of this study are available from the corresponding author upon reasonable request.
%

\bibliographystyle{MSP}



\end{document}